\documentclass[aps,prl,reprint,preprintnumbers]{revtex4-2}
\usepackage{blindtext}
\usepackage{array}
\usepackage{amsmath}
\usepackage{amsfonts}
\usepackage{amssymb}
\usepackage{amsthm}

\usepackage[english]{babel}
\usepackage[T1]{fontenc}
\usepackage{lmodern}

\usepackage{comment}
\usepackage{tikz}
\usetikzlibrary{decorations.markings}
\usepackage{bbm}

\usepackage{subcaption}  
\usepackage{caption}
\usepackage{graphicx}
\usepackage{hyperref}
\hypersetup{
   colorlinks=true,
   urlcolor=blue,    % color of external links
   linkcolor=black,  % color of toc, list of figs etc.
   citecolor=blue,   % color of links to bibliography1
}

\def \bea{\begin{eqnarray}}
	\def \eea{\end{eqnarray}}
\newcommand{\bean}{\begin{eqnarray*}}
	\newcommand{\eean}{\end{eqnarray*}}
\newcommand{\nn}{\nonumber \\}
\def\d{\partial}
\def\eref#1{(\ref{#1})}
\def\WH #1{\widehat{#1}}
\def\W #1{\widetilde{#1}}

\newcommand{\mi}{\mathrm{i}}
\newcommand{\md}{\mathrm{d}}
\def\Label#1{\label{#1}%
	\smash{\hbox to0pt{\raise1ex\hbox{\tiny[#1]}\hss}}}

\usepackage{color}  % red, green, blue, cyan, magenta, yellow, orange, violet,
\usepackage{xcolor}
\definecolor{ceil}{rgb}{0.57, 0.63,0.81}

\begin{document}

\begin{flushright}
{\tt USTC-ICTS/PCFT-25-37}
\end{flushright}
\title{
Symbolic Reduction of Multi-loop Feynman Integrals via Generating Functions
}
\author{Bo Feng$^{a,b,c,d}$}
\author{Xiang Li$^{e}$}
\author{Yuanche Liu $^{g}$}
\author{Yan-Qing Ma$^{e,f}$}
\author{Yang Zhang$^{g,d,f}$}
\date{August 2025}
\affiliation{
$^a$State Key Laboratory of Nuclear Physics and Technology, Institute of Quantum Matter, South China Normal University, Guangzhou 510006, China\\
$^b$Guangdong Basic Research Center of Excellence for Structure and Fundamental Interactions of Matter, Guangdong Provincial Key Laboratory of Nuclear Science, Guangzhou 510006, China\\
$^c$ Beijing Computational Science Research Center, Beijing 100084, China\\
$^d$ Peng Huanwu Center for Fundamental Theory, Hefei, Anhui, 230026, China\\
$^e$ School of Physics, Peking University, Beijing 100871, China\\
$^f$ Center for High Energy Physics, Peking University, Beijing 100871, China\\
$^g$ Interdisciplinary Center for Theoretical Study, University of Science and Technology of China, Hefei, Anhui 230026, China
}

\begin{abstract}
We introduce a novel, systematic method for the complete symbolic reduction of multi-loop Feynman integrals, leveraging the power of generating functions. The differential equations governing these generating functions naturally yield symbolic recurrence relations. We develop an efficient algorithm that utilizes these recurrences to reduce integrals to a minimal set of master integrals. This approach circumvents the exponential growth of traditional integration-by-parts relations, enabling the reduction of high-rank, multi-loop integrals critical for state-of-the-art calculations in perturbative quantum field theory.
\end{abstract}
\maketitle

\section{Introduction}
Loop-level Feynman diagrams play a central role in perturbative quantum field theories (QFTs). The current era of precision physics presents an urgent demand for the calculation of multi-loop amplitudes, which are crucial for collider phenomenology, gravitational wave predictions from black hole scattering, as well as formal aspects of QFT. In multi-loop computations, a key step is to reduce all Feynman integrals to a finite, linearly independent set consisting of master integrals. This step, known as Feynman integral reduction, is frequently the computationally intensive bottleneck of perturbative QFT calculation. Conventionally, integral reduction is carried out using the integration-by-parts (IBP) identities~\cite{Tkachov:1981wb,Chetyrkin:1981qh} via the linear reduction algorithm named after Laporta~\cite{Laporta:2000dsw}. However, for most cutting-edge multi-loop computations, the IBP method inevitably involves a vast linear system, which is prohibitively difficult to solve via linear reduction.

There are several public software packages for Feynman integral reduction, implementing the Laporta algorithm along with various modern techniques: 
{\sc AIR}~\cite{Anastasiou:2004vj}, 
{\sc FIRE}~\cite{Smirnov:2008iw,Smirnov:2013dia,Smirnov:2014hma,Smirnov:2019qkx}, 
{\sc LiteRed}~\cite{Lee:2012cn,Lee:2013mka}, 
{\sc Reduze}~\cite{Studerus:2009ye,vonManteuffel:2012np},
{\sc FiniteFlow}~\cite{Peraro:2016wsq},
{\sc Kira}~\cite{Maierhofer:2017gsa,Maierhofer:2018gpa,
Maierhofer:2019goc,Klappert:2020nbg,Lange:2025fba}, 
{\sc FireFly}~\cite{Klappert:2019emp,Klappert:2020aqs},
{\sc RaTracer}~\cite{Magerya:2022hvj},
{\sc Blade}~\cite{Guan:2019bcx,Liu:2021wks,Guan:2024byi}, 
and {\sc NeatIBP}~\cite{Wu:2023upw,Wu:2025aeg}.  
Notable advances that have boosted linear reduction include the finite-field method~\cite{vonManteuffel:2014ixa}, 
the application of computational algebraic geometry~\cite{Gluza:2010ws,Larsen:2015ped,Wu:2023upw,Wu:2025aeg}, 
and the block-triangular form~\cite{Guan:2019bcx,Liu:2021wks,Guan:2024byi}.  
An alternative framework is provided by intersection theory~\cite{Mastrolia:2018uzb,Chestnov:2022alh,%
Brunello:2024tqf}, formulating Feynman integral reduction in terms of twisted cohomology.  
Despite these developments, the capabilities of existing reduction methods are still insufficient to meet the demands of upcoming precision experiments, such as the High-Luminosity Large Hadron Collider (HL-LHC), the Laser Interferometer Space Antenna (LISA), the Taiji and TianQin projects.

The main bottleneck of integration-by-parts (IBP) reduction arises in multi-loop integrals with high-degree numerators or high-power propagators, where the number of IBP relations grows exponentially.
This issue becomes particularly severe in recent cutting-edge
multi-loop computations, such as the conservative $5$\,pm $2$\,SF black-hole
scattering~\cite{Bern:2025wyd, Driesse:2026qiz} which required $\sim 3\times 10^6$ core CPU hours
with {\sc Kira3}~\cite{Lange:2025fba}, and the three-loop five-point
Feynman integrals~\cite{Liu:2024ont, Chicherin:2025mvc} where the most
complicated family's IBP reduction used hundreds of unitarity cuts and
tens of thousands IBP relations for each cut. Tackling reduction problems at the three- and four-loop orders, or for two-loop multi-leg processes, necessitates new conceptual breakthroughs.

Recurrence rules~\cite{Smirnov:2005ky, Lee:2012cn} provide an efficient alternative to IBP reductions, particularly for the challenging task of reducing multi-loop integrals with high-rank numerators or elevated propagator powers. The implementations of recurrence rules include methods based on Gr{\"o}bner bases~\cite{Tarasov:1998nx, Gerdt:2004kt, Smirnov:2005ky, Smirnov:2006wh, Smirnov:2006tz} and heuristic algorithms~\cite{Lee:2012cn, Lee:2013mka}. However, a significant drawback was the computational cost associated with the recurrence rule derivation, which has limited their widespread adoption.

In this work, we present a novel and highly efficient method, based on generating functions~\cite{Feng:2022hyg,Guan:2023avw}, to systematically derive complete symbolic recurrence relations for the reduction of Feynman integrals. By expressing Feynman integrals in terms of generating functions, integration-by-parts (IBP) relations are translated into differential equations for these generating functions.  Using 
IBP relations and other conditions (such as integrability conditions), the reduction of Feynman integrals is then reformulated as reducing differential operators based on operator grading. This method will enable us to efficiently obtain complete symbolic reduction rules for arbitrary Feynman integrals with arbitrary indices.  

This paper is organized as follows. We first review the definition of generating functions for Feynman integral reduction. We then present our new systematic algorithm for obtaining symbolic reduction rules. As illustrations, we apply this method to three multi-loop reduction examples: the sunset diagram, the planar double-box diagram, and the non-planar double-box diagram.

During the preparation of this manuscript, we became aware of the recent paper~\cite{Smith:2025xes}, which presents an efficient method for generating Feynman integral reduction rules using syzygies. In this work, we emphasize that our method uses only fundamental IBP relations and does not rely on syzygies.

\section{Generating Functions for Feynman integrals}
A Feynman integral family with $L$ loops and $E$ independent legs can be represented as
\bea
	I(\vec{\nu})=\int\prod_{i=1}^{L}\frac{\md^{D}\ell_i}{\mi\pi^{D/2}}
	\frac{\mathcal{D}_{K+1}^{-\nu_{K+1}}\cdots \mathcal{D}_N^{-\nu_N}}{\mathcal{D}_1^{\nu_{1}}\cdots \mathcal{D}_K^{\nu_{K}}},~~~\label{Idea-1-1}
\eea
where $D = 4-2\epsilon$ represents the spacetime dimension, $\ell_i$ are the loop momenta, $\mathcal{D}_1,\ldots,\mathcal{D}_K$ are the inverse propagators, $\mathcal{D}_{K+1},\ldots,\mathcal{D}_N$ are the irreducible scalar products (ISPs) introduced to form a complete basis and $N={L(L+1)\over 2}+LE$. 
In \eref{Idea-1-1},  $\nu_1,\ldots,\nu_K$ can be arbitrary integers (i.e, positive, negative or zero), while $\nu_{K+1},\ldots,\nu_N$ can only be non-positive integers. An integral is said to belong to the top-sector of this family if all its propagator indices are positive, i.e., $\nu_1, \ldots, \nu_K > 0$. 

To organize all integrals within a sector, we introduce a generating function defined as:
\bea G_{\vec{\mu}}(\eta)= \int\prod_{i=1}^{L}\frac{\md^{D}\ell_i}{\mi\pi^{D/2}} {e^{\sum_{j=1}^N (1-\mu_j)\eta_j s_0^{-1}\mathcal{D}_j}
 \over \prod_{i=1}^N(\mathcal{D}_i-s_0\eta_i)^{\mu_i} },~~~\label{Idea-1-5}\eea
where $\vec{\mu}=(\mu_1,...,\mu_N)$ consists of $\mu_i=0,1$ and $s_0$ is an auxiliary scale that make $\eta_i$ dimensionless. For top-sector, we set $\mu_1,\cdots,\mu_K=1$ and $\mu_{K+1},\cdots,\mu_N=0$ and $G_{\vec{\mu}}$ can be expanded as:
\bea G_{\vec{\mu}}(\vec{\eta})&= &\sum_{\vec{n}\geq 0} \vec{\eta}^{\vec{n}} F_{\vec{n}},~~~\label{Idea-1-5-2} \eea
with
\bea
F_{\vec{n}} &= & \left(\prod_{j=K+1}^N {1\over n_j!}\right) s_0^{\sum_{i=1}^K n_i-\sum_{j=K+1}^N n_j} \nn
&\times &I(...,n_K+1, -n_{K+1},...,-n_N) ~. \eea
The formalism \eref{Idea-1-5-2} has mapped Feynman integrals to the expansion coefficients of generating function, labeled by the lattice points $\vec{n}$. We define the {\bf degree} of $F$ as $\sum_{i=1}^N n_i$. Consequently, the reduction problem of Feynman integrals is transformed into finding relations among these coefficients, which can be derived from differential equations(DEs) satisfied by the generating function.

 The  differential equations for generating function have the form
\bea \sum_t c_t \hat{O}_t G_{\vec{\mu}}(\vec{\eta})= B~,~~~\hat{O}_t=\prod_{i=1}^N \eta_i^{(\vec{a}_t)_i} {\d^{(\vec{b}_t)_i}\over 
	\d\eta_i^{(\vec{b}_t)_i}~},~\label{lesson-1-1-1}\eea
where $B$ is the function of generating functions of subsectors,  $\hat{O}_t$'s are the operator writing into the standard form   and $c_t$ are coefficients. In the standard form, the operator is uniquely fixed by a pair of vectors $(\vec{a},\vec{b})$, which is called the {\bf finer index of operator}. By %substituting the expansion \eref{Idea-1-5-2} to equation \eref{lesson-1-1-1} and 
extracting the coefficient of $\vec{\eta}^{\vec{n}}$, we obtain
\bea \sum_t c_t \left(\prod_{j=1}^N\prod_{p=1}^{(\vec{b}_t)_j}(n_j-(\vec{a})_j+p)\right) F_{\vec{n}+\vec{b}_t-\vec{a}_t}= \W B,~~~~\label{lesson-1-1-1a}\eea
where for simplicity, we have made the convention that $F_{\vec{n}\not\geq 0}=0$ \footnote{$\vec{a}\geq 0$, means that every component of the vector is non-negative integer and  $\vec{a}\not\geq 0$ means that some components of the vector will be negative integer.}.
We call equation \eref{lesson-1-1-1a} as the {\bf recurrence relation} derived from the DE \eref{lesson-1-1-1}.

From \eref{lesson-1-1-1a}, one can see the role of operators in the recurrence relation is largely fixed by the 
 combination $\vec{o} \equiv \vec{b} - \vec{a}$, which is called the {\bf index} of the operator. Similarly, we define the {\bf degree} of an operator as $|\vec{o}|\equiv \sum_{i=1}^N (\vec{o})_i$, which relates to the degree of lattice points via \eref{lesson-1-1-1a}. Furthermore, we call an equation of degree $M$, if it contains operators with the highest degree $M$. For an operator with more than one terms of the highest degree, the unique leading term would be defined in the next section for the Module II.

The mapping between  \eref{lesson-1-1-1} and  \eref{lesson-1-1-1a} enables  the reduction of Feynman integrals, i.e., obtaining 
coefficients of higher-degree lattice points in \eref{Idea-1-5-2}, becomes a problem to reduce a higher-degree operator as a linear combination of lower-degree operators using the DEs. The primary objective of this work is to develop a systematic algorithm to solve this task.

In the classic work \cite{Smirnov:2005ky}, IBP reduction was reformulated as an operator reduction problem and solved via Gr\"obner basis. In this work, we emphasize that Gr\"obner basis is not used for reducing operators.

We comment on the operators. First, there are special operators with 
finer index $(\vec{a},\vec{a})$. %Since their index $\vec{o}=\vec{0}$, we will call them 
which are called  {\bf zero-index operators} since $\vec{o}=\vec{0}$.\footnote{Later, a zero-index operator  could also mean a linear combination of zero-index operators}. An important fact is that 
any two  zero-index operators  commute. Second, there are many operators with the same index. All of them can be written into the form $\WH O_0 \WH O^F_{\vec{o}}$, where $\WH O_0$ is a zero-index operator and $\WH O^F_{\vec{o}}\equiv \vec{\eta}^{\vec{o}_{-} }{\d^{\vec{o}_+}\over \d \vec{\eta}^{\vec{o}_+}}$. The  $\vec{o}_\pm$ are   the positive and negative components of $\vec{o}$ such that $\vec{o}=\vec{o}_+-\vec{o}_{-}$. Thus, we introduce a useful concept: an operator $\WH Q_b$ is called {\bf the descendant} of an operator $\WH Q_a$ if there is another operator %(or the combination of operators) 
$\WH O_1$, such that
\bea \WH O_0\WH Q_b= \WH O_1 \WH Q_a,~~~~\label{Index-zero-des-1-1}\eea
where $\WH O_0$ is a zero-index operator.

\section{Symbolic Reduction Rules }

The DEs  can be categorized as the so-called T1-type or T2-type.  An equation is the T1-type if all highest degree operators therein have the same  operator index. Otherwise it is a T2-type equation. We would reduce the highest operators having a given index $\vec{o}$. 

For a relation to be applicable for all choices $\vec{n}\geq 0$ in \eref{lesson-1-1-1a}, it is required that $\vec{o}\geq 0$. Depending whether there is one or 
multiple operators for the index $\vec{o}$, the reduction can be divided into the A-type or B-Type. More explicitly, %the symbolic reduction rules 
%obtained using our algorithm will be following forms:
%
\bea \text{T1A-type}: &~~~~& {\d^{\vec{b}}\over \d \eta^{\vec{b}}} G_{\vec{\mu}}(\eta)= R_{\vec{b}}+B~~~~\label{T1A-red} \\
\text{T1B-type}: &~~~~& \WH O_0 {\d^{\vec{b}}\over \d \eta^{\vec{b}}} G_{\vec{\mu}}(\eta)= R_{\vec{b}}+B~~~~\label{T1B-red}\\
\text{T2A-type}: &~~~~& {\d^{\vec{b}}\over \d \eta^{\vec{b}}} G_{\vec{\mu}}(\eta)=Q_{\vec{b}} + R_{\vec{b}}+B~~~~\label{T2A-red} \\
\text{T2B-type}: &~~~~& \WH O_0 {\d^{\vec{b}}\over \d \eta^{\vec{b}}}=Q_{\vec{b}} + R_{\vec{b}}+B~~~~\label{T2B-red} \eea
where $\WH O_0=\left[c_0+\sum_i c_i \eta^{\vec{a}_i} {\d^{\vec{a}_i}\over \d \eta^{\vec{a}_i}}\right]$ is a zero-index operator. 
The $Q$ part contains operators with the same degree as the left hand side, while the $R$ part contains only operators with lower degree. Finally $B$ contains contributions from sub-sectors. Importantly, we further require $\WH O_0$ containing identity operator for 
 \eref{T1B-red} and \eref{T2B-red} (i.e., $c_0\neq 0$).

Our novel algorithm  can be summarized by the diagram with three modules, or {\it a golden triangle},

\begin{center}
    {\tiny
        \begin{tikzpicture}[node distance = 2 cm, auto]

            % Boxes
            \node[draw, black, align=center] at (-3.5,0) {I: Generating \\Equations};
            \node[draw, black, align=center] at (0,4.5) {II: Solving Equations \\to Get Reduction Rules};
            \node[draw, red, align=center] at (3.5,0) {III: Reduction rules \\Complete?};

            % Arrows and labels
            \draw[<-][very thick] (-2,0) -- (2,0);
            \draw[<-][thick] (-3.5,-0.4) -- (-3.5,-1.5);
            \node[below] at (-3.5,-1.7) {fundamental IBPs};

            \draw[->][very thick] (-3.5,0.5) -- (-0.5,4);
            \draw[->][very thick] (0.5,4) -- (3.5,0.5);

            \draw[->][thick, red] (3.5,-0.4) -- (3.5,-1.5);
            \node[below, red] at (3.5,-1.7) {Finish};
            \node[right, red] at (3.5,-0.7) {Yes};
            \node[above] at (0,0) {No};

        \end{tikzpicture}
    }~~~~~\label{Alg-dia}
\end{center}

We present a brief explanation for each module:
\begin{itemize}
	\item {\bf Module I:} Input DEs come from two different origins: (a) the initial seeds of DEs from fundamental IBP relations; (b)  all others from the action of ${\d\over \d \eta_i}$ on reduction rules. %found in the  previous solving module.
	
	 After the action of ${\d\over \d \eta_i}$, a crucial step is to use all known reduction rules to simplify it. For the highest degree operator $\WH Q_b$, if it is the descendant of the reduction rule $\WH Q_a$ %(i.e., there is a relation like \eref{Index-zero-des-1-1}), 
     then in the following combination according to \eref{Index-zero-des-1-1},
     \begin{equation}
         \WH Q_0 {\rm Eq}- \WH Q_1 {\rm Rule}_{\WH Q_a}
     \end{equation}
     eliminate the highest degree part. 
	 
	% We use this method to simplify all highest degree operators in the equation 
     Repeat this procedure until there is no  highest degree operators to be simplified.  If all highest operators have been reduced, we move to the next highest degree operators and repeat.  
	 
	% For the equation obtained with the action ${\d\over \d\eta_i}$, only when it can be simplified by at least another reduction rule, it can be the taken as the new equation to be solved. 
	
	\item {\bf Module II:} Classify equations from the module I as different subsets according to  degrees. %(i.e., the highest degree of operators in the Eq's). 
    Then we start solving the set with highest degree and then deal with lower-degree sets, until we reach the degree zero. For each subset, the highest degree operators can be reduced by linear algebra methods, say, Gaussian elimination. 
    
    When solving the T2-type equations, which have highest degree operators with different operator indices, we must carefully define a global priority ordering. The core policy is to prevent the formation of loops in the reduction rules, which is achieved by ensuring a net decrease in at least one index throughout the recursive process, which guarantees the eventual termination of the reduction chain.

    After solving the subset, new  reduction rules are implemented to simplify previous reduction rules and remaining unsolved equations.  This step is called {\bf update}. 
	
	\item {\bf Module III:} With a reduction rule for the operator with index $\vec{o}$, the lattice points that can be reduced by this rule are in the set,
	\bea {\cal S}_{\vec{o}} 
	& = & \{ \vec{m}\in \mathbb Z^N| \vec{m}\geq  0~~ \& ~~ \vec{m}- \vec{o}\geq 0\}\,,~~~~\label{red-set}\eea 
	and the irreducible lattice points are in the set,
	\bea {\cal U}_{\vec{o}}=\{ \vec{m}\in \mathbb Z^N| \vec{m}\geq  0~~ \& ~~ \vec{m}- \vec{o}\not\geq 0\}\,,~~~~\label{unred-set}\eea
	Thus, with all reduction rules, the set of irreducible lattice points at this point is  ${\cal U}_{total}\equiv \cap_{i=1}^P {\cal U}_{\vec{o}_i}$. Then count the number of lattice points in the set. if the number  equals  the number of master integrals in this sector,   reduction rules are complete. Otherwise, start a new round of computations with three modules.

\end{itemize}

Our framework offers distinct advantages over existing methods. Compared to methods based on Gr{\"o}bner bases~\cite{Tarasov:1998nx, Gerdt:2004kt, Smirnov:2005ky, Smirnov:2006wh, Smirnov:2006tz}, our approach circumvents complex non-commutative algebra. Instead, it systematically reduces the derivation of reduction rules to a more tractable linear algebra problem. Compared to heuristic algorithms~\cite{Lee:2012cn, Lee:2013mka}, our approach provides a general and reusable solution with a clear termination condition. This contrasts sharply with the ad-hoc nature of heuristic strategies, which lack guarantees of completeness and often fail when faced with exceptionally complex topologies. %\color{red} The problem is our examples not contain exceptionally complex topologies.\color{black}

Furthermore, our method streamlines previous generating function formalisms~\cite{Guan:2023avw}. By employing a single generating function per sector, we avoid the expensive cost of constructing closed differential equations for a large number of master generating functions, which is achieved by Laporta's algorithm in Ref.~\cite{Guan:2023avw}. And by introducing a novel simplification technique on the unreduced set, we significantly accelerate the generation of recurrence rules and descendant equations, making the entire reduction process more direct and computationally tractable.

Following this general framework, we will present several examples to demonstrate the algorithm of finding complete symbolic reduction rules in the following section. Our analysis will focus exclusively on top-sector integrals, because any integral in sub-sector can be treated as a top-sector integral within a simpler sub-family, defined by the propagators with non-positive powers.

%%%%%%%%%%%%%%%%%%%%%%%%%%%
\section{Example 1: the top sector of sunset diagrams}
%%%%%%%%%%%%%%%%%%%%%%%%%%

This is a simplest example to clearly show our algorithm. The sunset diagram has the propagators and ISPs, 
\bea \mathcal{D}_1 &= &(\ell_1^2-m_1^2),~~~ \mathcal{D}_2=(\ell_2^2-m_2^2),\nn \mathcal{D}_3 &= &((\ell_1+\ell_2-K)^2-m_3^2),\nn
\mathcal{D}_4&= &\ell_1\cdot K,~~~~ \mathcal{D}_5=\ell_2\cdot K.~~~\label{sunset-1-2}\eea
The details of computations can be found in the Supplemental Material, here we will just present the outline of the whole procedure. 

{\bf The first round of reduction:} The initial seed for top-sector $G_{11100}=G_{111}$ is six degree two DE's from 
fundamental IBP relations. Using them, we find $6$ T1A-type reduction rules for  degree-2 operators. ${\d\over \d \eta_a} {\d\over \d \eta_i}$, $a=4,5; i=1,2,3$. Using them the irreducible lattice points are  $(0, 0, 0, n_4, n_5)$ and $(n_1, n_2, n_3, 0, 0)$. 

{\bf The second round of reduction:} Now generate new equations
by applying ${\d\over \d\eta_i}$ over the $6$ reduction rules above. %After simplification, 
Then we get $9$ nontrivial DE's, out of which only $8$ are independent.
The degree of the DE's depends on the mass configuration.

For the massless case, there are $3$ degree-two DE's and $5$ degree-one DE's. Using them, we can find T1B-type reduction rules %for  ${\d\over \d\eta_a}, a=4,5$ 
and T2B-type reduction rules for  %${\d\over \d\eta_i}, i=1,2$ using ${\d\over \d\eta_3}$. 
Thus the irreducible lattice points are just $(0, 0, n_3, 0, 0)$.

For the massive case, there are $5$ degree-two DE's and $3$ degree-one DE's. Using them, we  find T2B-type reduction rules for  %${\d\over \d\eta_a}, a=4,5$ using ${\d\over \d\eta_i}, i=1,2,3$, 
and $5$ T2A-type reduction rules.
%${\d^2\over\d\eta_i^2}, 
%i=1,2$ and ${\d^2\over\d\eta_i\d\eta_j}, i<j, i,j=1,2,3$. 
The irreducible points consist of the set $(0,0,n_3,0,0)$ and points  $(1,0,0,0,0)$, $(0,1,0,0,0)$.

{\bf The third round of computation:} Now we act ${\d\over \d\eta_3}$ on the reduction rule of, for example, ${\d\over \d\eta_4}$. For massless case,
we can find a T1B-type reduction rule. %for ${\d\over \d\eta_3}$. 
Thus the only irreducible point corresponds the master integrals $(0,0,0,0,0)$. For the massive case, we can find T1B-type reduction rule for ${\d^2\over \d\eta_3^2}$. 
Thus four points not reduced, i.e., $(0,0,0,0,0), (1,0,0,0,0),(0,1,0,0,0), (0,0,1, 0,0)$,  correspond to the master integrals for massive sunset in the top sector.

%%%%%%%%%%%%%%%%%%%%%%%%%%%
\section{Example 2: the top sector of nonplanar double box diagrams}
%%%%%%%%%%%%%%%%%%%%%%%%%%

This is a nontrivial example. A nonplanar two-loop integral can be reduced in an elegant way. We choose a complete set of Lorentz scalars as
\begin{align}
	\begin{split}
		&\mathcal{D}_1={\ell_1}^2,~~ \mathcal{D}_2=(\ell_1-k_1)^2,~~~\mathcal{D}_3={(\ell_1-k_1-k_2)}^2,\\
		&\mathcal{D}_4= {\ell_2}^2,~\mathcal{D}_5={(\ell_2+k_4)}^2,~\mathcal{D}_6= {(\ell_2-\ell_1+k_1+k_2+k_4)}^2,\\
		&\mathcal{D}_7={(\ell_2-\ell_1)}^2,~~~ \mathcal{D}_8 = \ell_1 \cdot k_4,~~~ \mathcal{D}_9 = \ell_2 \cdot k_1,~~~\label{nonplanerBox-1-1-1}
	\end{split}
\end{align}
where the last two are ISPs. The kinematics specified as $k_i^2=0$, $i=1,2,4$ and $k_1\cdot  k_2=s/2$, $k_2 \cdot k_3=t/2$ and $k_1\cdot  k_3=-(s+t)/2$.

{\bf The first round computation:} The initial seeds are the $10$ DE's from fundamental IBP relations. From them, we can solve $2$ T1A-type reduction rules for degree-one operators. %${\d\over \d\eta_i}, i=1,3$, 
$6$ T1A-type reduction rules for degree-two operators. %${\d\over \d\eta_8}{\d\over \d\eta_j}, j=2,4,5,6,7$ and %${\d\over \d\eta_9}{\d\over \d\eta_2}$. 
There are also two T2A-type reduction rules. %One is the reduction rule of degree-2 operator ${\d\over \d\eta_9}{\d\over \d\eta_6}$, which depends on ${\d\over \d\eta_9}{\d\over \d\eta_7}$. 
Using these reduction rules,  irreducible lattice points are 
\bea {\cal U}_{11} &= &(0,n_2, 0,n_4, n_5, n_6, n_7, 0, 0),\nn
{\cal U}_{12} & = & (0,0, 0,0, n_5, 0, n_7, 0, n_9),\nn
{\cal U}_2 &= & (0,0, 0,0, 0, 0, 0, n_8, n_9). ~~~~~\label{np-un-point-2}\eea

{\bf The second round computation:} Since the indices for $n_1, n_3$ have been fully reduced, consider the action ${\d\over \d\eta_i}, i\neq 1,3$ on above $10$ reduction rules. Among $70$ equations, there are $16$ nontrivial DE's. Among them, there are $12$ degree-two DE's and four degree-one DE's.  We can solve $10$ T1A-type reduction rules for degree-two operators.
  %${\d\over \d\eta_2}{\d\over \d\eta_4}$,  ${\d\over \d\eta_2}{\d\over \d\eta_5}$,  ${\d\over \d\eta_2}{\d\over \d\eta_6}$,  ${\d\over \d\eta_2}{\d\over \d\eta_7}$, ${\d\over \d\eta_4}{\d\over \d\eta_5}$, ${\d\over \d\eta_4}{\d\over \d\eta_7}$,  ${\d\over \d\eta_5}{\d\over \d\eta_9}$, ${\d\over \d\eta_5}{\d\over \d\eta_6}$,  ${\d\over \d\eta_6}{\d\over \d\eta_7}$,  ${\d\over \d\eta_7}{\d\over \d\eta_9}$. 
  A T2A-type reduction rule, %for, example, for 
  %${\d\over \d\eta_5}{\d\over \d\eta_7}$ by ${\d\over \d\eta_4}{\d\over \d\eta_6}$ 
  and a T2B-reduction rule are obtained. %for, example, for ${\d\over \d\eta_8}{\d\over \d\eta_9}$ by ${\d^2\over \d\eta_8^2} $. 
  From the degree-one DE's, three T2B-type reduction rules can be solved. %for example, using ${\d\over \d\eta_4}$ to solve ${\d\over \d\eta_6}$, ${\d\over \d\eta_5}$ to solve ${\d\over \d\eta_7}$ and ${\d\over \d\eta_4}$ and ${\d\over \d\eta_5}$ to solve ${\d\over \d\eta_2}$.  
  Now the irreducible points are $(0,0,0,n_4,n_5,0,0,0,0)$, $(0,0,0,0,0,0,0,n_8,0)$ and $(0,0,0,0,0,0,0,0,n_9)$.

{\bf The third round computation:}  Since ${\d\over \d\eta_2}, {\d\over \d\eta_6}, {\d\over \d\eta_7}$ have been fully reduced, we consider the action of ${\d\over \d\eta_i}, i=4,5,8,9$ only. To generate fewer equations, we act only on three degree-one reduction rules found in previous round. Using these $12$ DE's, we can solve $3$ T1B-type reduction rules for degree-2 operators ${\d\over \d\eta_4}  {\d\over \d\eta_6}$, $ {\d^2\over \d\eta_4^2}$ and  ${\d^2\over \d\eta_5^2}$. Furthermore, we can solve ${\d\over \d \eta_9}$ %by ${\d\over \d \eta_8}$ and ${\d\over \d \eta_5}$ 
, ${\d\over \d \eta_8}$,
%by ${\d\over \d \eta_4}$, 
and ${\d\over \d \eta_5}$. %by ${\d\over \d \eta_4}$. 
Using these reduction rules, reduce $n_5, n_8, n_9$ are reduced to zero, and by the T1B-reduction rule of  ${\d^2\over \d\eta_4^2}$,  $n_4$ to $0,1$ are reduced to zero. So thus on the top sector, all integrals are reduced to two master integrals.

\section{Conclusion and Outlook}
In this work, we introduced a novel systematic approach to derive complete symbolic reduction rules for multi-loop Feynman integrals with generating functions. By the DEs for these generating functions, we obtain recurrence relations that enable the efficient reduction without the exponential proliferation of IBP identities in traditional methods.  In other words, the IBP relations are solved {\it analytically.} Through explicit examples, including the sunset diagrams (both massive and massless), two-loop double-box, and two-loop non-planar integrals, we demonstrate the algorithm's effectiveness in  complex cases with high-degree numerators and high-power propagators. %This method addresses key bottlenecks in perturbative QFT computations.

With the proof of concepts in the note, the most important thing is to automate the aforementioned algorithm. In a longer companion paper, more details of the algorithm  will be presented and cutting-edge examples like the reduction of two-loop five-point massless and massive Feynman integrals  for collider phenomenology.

The combination of the idea in the paper and other reduction methods, as well as the application on more general input other than Feynman integrals, like energy-energy correlator and cosmological correlator, are also worth pursuing.   

\begin{acknowledgments}
We thank Mingxing Luo for many helpfull discussions. Bo Feng is supported by the National Natural Science Foundation of China (NSFC) through Grants No. 12535003, No.11935013, No.11947301, No.12047502.  Xiang Li and Yan-Qing Ma are supported by NSFC through Grant No. 12325503.  Yuanche Liu is  supported by NSFC through Grant No. 124B1014, and Yang Zhang is supported by NSFC through Grant No. 12575078 and  12247103.
\end{acknowledgments}
\bibliography{main}

%apsrev4-2.bst 2019-01-14 (MD) hand-edited version of apsrev4-1.bst
%Control: key (0)
%Control: author (8) initials jnrlst
%Control: editor formatted (1) identically to author
%Control: production of article title (0) allowed
%Control: page (0) single
%Control: year (1) truncated
%Control: production of eprint (0) enabled
\begin{thebibliography}{46}%
\makeatletter
\providecommand \@ifxundefined [1]{%
 \@ifx{#1\undefined}
}%
\providecommand \@ifnum [1]{%
 \ifnum #1\expandafter \@firstoftwo
 \else \expandafter \@secondoftwo
 \fi
}%
\providecommand \@ifx [1]{%
 \ifx #1\expandafter \@firstoftwo
 \else \expandafter \@secondoftwo
 \fi
}%
\providecommand \natexlab [1]{#1}%
\providecommand \enquote  [1]{``#1''}%
\providecommand \bibnamefont  [1]{#1}%
\providecommand \bibfnamefont [1]{#1}%
\providecommand \citenamefont [1]{#1}%
\providecommand \href@noop [0]{\@secondoftwo}%
\providecommand \href [0]{\begingroup \@sanitize@url \@href}%
\providecommand \@href[1]{\@@startlink{#1}\@@href}%
\providecommand \@@href[1]{\endgroup#1\@@endlink}%
\providecommand \@sanitize@url [0]{\catcode `\\12\catcode `\$12\catcode `\&12\catcode `\#12\catcode `\^12\catcode `\_12\catcode `\%12\relax}%
\providecommand \@@startlink[1]{}%
\providecommand \@@endlink[0]{}%
\providecommand \url  [0]{\begingroup\@sanitize@url \@url }%
\providecommand \@url [1]{\endgroup\@href {#1}{\urlprefix }}%
\providecommand \urlprefix  [0]{URL }%
\providecommand \Eprint [0]{\href }%
\providecommand \doibase [0]{https://doi.org/}%
\providecommand \selectlanguage [0]{\@gobble}%
\providecommand \bibinfo  [0]{\@secondoftwo}%
\providecommand \bibfield  [0]{\@secondoftwo}%
\providecommand \translation [1]{[#1]}%
\providecommand \BibitemOpen [0]{}%
\providecommand \bibitemStop [0]{}%
\providecommand \bibitemNoStop [0]{.\EOS\space}%
\providecommand \EOS [0]{\spacefactor3000\relax}%
\providecommand \BibitemShut  [1]{\csname bibitem#1\endcsname}%
\let\auto@bib@innerbib\@empty
%</preamble>
\bibitem [{\citenamefont {Tkachov}(1981)}]{Tkachov:1981wb}%
  \BibitemOpen
  \bibfield  {author} {\bibinfo {author} {\bibfnamefont {F.~V.}\ \bibnamefont {Tkachov}},\ }\bibfield  {title} {\bibinfo {title} {{A Theorem on Analytical Calculability of Four Loop Renormalization Group Functions}},\ }\href {https://doi.org/10.1016/0370-2693(81)90288-4} {\bibfield  {journal} {\bibinfo  {journal} {Phys. Lett. B}\ }\textbf {\bibinfo {volume} {100}},\ \bibinfo {pages} {65} (\bibinfo {year} {1981})}\BibitemShut {NoStop}%
\bibitem [{\citenamefont {Chetyrkin}\ and\ \citenamefont {Tkachov}(1981)}]{Chetyrkin:1981qh}%
  \BibitemOpen
  \bibfield  {author} {\bibinfo {author} {\bibfnamefont {K.~G.}\ \bibnamefont {Chetyrkin}}\ and\ \bibinfo {author} {\bibfnamefont {F.~V.}\ \bibnamefont {Tkachov}},\ }\bibfield  {title} {\bibinfo {title} {{Integration by Parts: The Algorithm to Calculate beta Functions in 4 Loops}},\ }\href {https://doi.org/10.1016/0550-3213(81)90199-1} {\bibfield  {journal} {\bibinfo  {journal} {Nucl. Phys. B}\ }\textbf {\bibinfo {volume} {192}},\ \bibinfo {pages} {159} (\bibinfo {year} {1981})}\BibitemShut {NoStop}%
\bibitem [{\citenamefont {Laporta}(2000)}]{Laporta:2000dsw}%
  \BibitemOpen
  \bibfield  {author} {\bibinfo {author} {\bibfnamefont {S.}~\bibnamefont {Laporta}},\ }\bibfield  {title} {\bibinfo {title} {{High-precision calculation of multiloop Feynman integrals by difference equations}},\ }\href {https://doi.org/10.1142/S0217751X00002159} {\bibfield  {journal} {\bibinfo  {journal} {Int. J. Mod. Phys. A}\ }\textbf {\bibinfo {volume} {15}},\ \bibinfo {pages} {5087} (\bibinfo {year} {2000})},\ \Eprint {https://arxiv.org/abs/hep-ph/0102033} {arXiv:hep-ph/0102033} \BibitemShut {NoStop}%
\bibitem [{\citenamefont {Anastasiou}\ and\ \citenamefont {Lazopoulos}(2004)}]{Anastasiou:2004vj}%
  \BibitemOpen
  \bibfield  {author} {\bibinfo {author} {\bibfnamefont {C.}~\bibnamefont {Anastasiou}}\ and\ \bibinfo {author} {\bibfnamefont {A.}~\bibnamefont {Lazopoulos}},\ }\bibfield  {title} {\bibinfo {title} {{Automatic integral reduction for higher order perturbative calculations}},\ }\href {https://doi.org/10.1088/1126-6708/2004/07/046} {\bibfield  {journal} {\bibinfo  {journal} {JHEP}\ }\textbf {\bibinfo {volume} {07}},\ \bibinfo {pages} {046}},\ \Eprint {https://arxiv.org/abs/hep-ph/0404258} {arXiv:hep-ph/0404258} \BibitemShut {NoStop}%
\bibitem [{\citenamefont {Smirnov}(2008)}]{Smirnov:2008iw}%
  \BibitemOpen
  \bibfield  {author} {\bibinfo {author} {\bibfnamefont {A.~V.}\ \bibnamefont {Smirnov}},\ }\bibfield  {title} {\bibinfo {title} {{Algorithm FIRE -- Feynman Integral REduction}},\ }\href {https://doi.org/10.1088/1126-6708/2008/10/107} {\bibfield  {journal} {\bibinfo  {journal} {JHEP}\ }\textbf {\bibinfo {volume} {10}},\ \bibinfo {pages} {107}},\ \Eprint {https://arxiv.org/abs/0807.3243} {arXiv:0807.3243 [hep-ph]} \BibitemShut {NoStop}%
\bibitem [{\citenamefont {Smirnov}\ and\ \citenamefont {Smirnov}(2013)}]{Smirnov:2013dia}%
  \BibitemOpen
  \bibfield  {author} {\bibinfo {author} {\bibfnamefont {A.~V.}\ \bibnamefont {Smirnov}}\ and\ \bibinfo {author} {\bibfnamefont {V.~A.}\ \bibnamefont {Smirnov}},\ }\bibfield  {title} {\bibinfo {title} {{FIRE4, LiteRed and accompanying tools to solve integration by parts relations}},\ }\href {https://doi.org/10.1016/j.cpc.2013.06.016} {\bibfield  {journal} {\bibinfo  {journal} {Comput. Phys. Commun.}\ }\textbf {\bibinfo {volume} {184}},\ \bibinfo {pages} {2820} (\bibinfo {year} {2013})},\ \Eprint {https://arxiv.org/abs/1302.5885} {arXiv:1302.5885 [hep-ph]} \BibitemShut {NoStop}%
\bibitem [{\citenamefont {Smirnov}(2015)}]{Smirnov:2014hma}%
  \BibitemOpen
  \bibfield  {author} {\bibinfo {author} {\bibfnamefont {A.~V.}\ \bibnamefont {Smirnov}},\ }\bibfield  {title} {\bibinfo {title} {{FIRE5: A C++ implementation of Feynman Integral REduction}},\ }\href {https://doi.org/10.1016/j.cpc.2014.11.024} {\bibfield  {journal} {\bibinfo  {journal} {Comput. Phys. Commun.}\ }\textbf {\bibinfo {volume} {189}},\ \bibinfo {pages} {182} (\bibinfo {year} {2015})},\ \Eprint {https://arxiv.org/abs/1408.2372} {arXiv:1408.2372 [hep-ph]} \BibitemShut {NoStop}%
\bibitem [{\citenamefont {Smirnov}\ and\ \citenamefont {Chukharev}(2020)}]{Smirnov:2019qkx}%
  \BibitemOpen
  \bibfield  {author} {\bibinfo {author} {\bibfnamefont {A.~V.}\ \bibnamefont {Smirnov}}\ and\ \bibinfo {author} {\bibfnamefont {F.~S.}\ \bibnamefont {Chukharev}},\ }\bibfield  {title} {\bibinfo {title} {{FIRE6: Feynman Integral REduction with modular arithmetic}},\ }\href {https://doi.org/10.1016/j.cpc.2019.106877} {\bibfield  {journal} {\bibinfo  {journal} {Comput. Phys. Commun.}\ }\textbf {\bibinfo {volume} {247}},\ \bibinfo {pages} {106877} (\bibinfo {year} {2020})},\ \Eprint {https://arxiv.org/abs/1901.07808} {arXiv:1901.07808 [hep-ph]} \BibitemShut {NoStop}%
\bibitem [{\citenamefont {Lee}(2012)}]{Lee:2012cn}%
  \BibitemOpen
  \bibfield  {author} {\bibinfo {author} {\bibfnamefont {R.~N.}\ \bibnamefont {Lee}},\ }\bibfield  {title} {\bibinfo {title} {{Presenting LiteRed: a tool for the Loop InTEgrals REDuction}},\ }\href@noop {} {\  (\bibinfo {year} {2012})},\ \Eprint {https://arxiv.org/abs/1212.2685} {arXiv:1212.2685 [hep-ph]} \BibitemShut {NoStop}%
\bibitem [{\citenamefont {Lee}(2014)}]{Lee:2013mka}%
  \BibitemOpen
  \bibfield  {author} {\bibinfo {author} {\bibfnamefont {R.~N.}\ \bibnamefont {Lee}},\ }\bibfield  {title} {\bibinfo {title} {{LiteRed 1.4: a powerful tool for reduction of multiloop integrals}},\ }\href {https://doi.org/10.1088/1742-6596/523/1/012059} {\bibfield  {journal} {\bibinfo  {journal} {J. Phys. Conf. Ser.}\ }\textbf {\bibinfo {volume} {523}},\ \bibinfo {pages} {012059} (\bibinfo {year} {2014})},\ \Eprint {https://arxiv.org/abs/1310.1145} {arXiv:1310.1145 [hep-ph]} \BibitemShut {NoStop}%
\bibitem [{\citenamefont {Studerus}(2010)}]{Studerus:2009ye}%
  \BibitemOpen
  \bibfield  {author} {\bibinfo {author} {\bibfnamefont {C.}~\bibnamefont {Studerus}},\ }\bibfield  {title} {\bibinfo {title} {{Reduze~{\textendash} Feynman integral reduction in C++}},\ }\href {https://doi.org/10.1016/j.cpc.2010.03.012} {\bibfield  {journal} {\bibinfo  {journal} {Comput. Phys. Commun.}\ }\textbf {\bibinfo {volume} {181}},\ \bibinfo {pages} {1293} (\bibinfo {year} {2010})},\ \Eprint {https://arxiv.org/abs/0912.2546} {arXiv:0912.2546 [physics.comp-ph]} \BibitemShut {NoStop}%
\bibitem [{\citenamefont {von Manteuffel}\ and\ \citenamefont {Studerus}(2012)}]{vonManteuffel:2012np}%
  \BibitemOpen
  \bibfield  {author} {\bibinfo {author} {\bibfnamefont {A.}~\bibnamefont {von Manteuffel}}\ and\ \bibinfo {author} {\bibfnamefont {C.}~\bibnamefont {Studerus}},\ }\bibfield  {title} {\bibinfo {title} {{Reduze 2 - Distributed Feynman Integral Reduction}},\ }\href@noop {} {\  (\bibinfo {year} {2012})},\ \Eprint {https://arxiv.org/abs/1201.4330} {arXiv:1201.4330 [hep-ph]} \BibitemShut {NoStop}%
\bibitem [{\citenamefont {Peraro}(2016)}]{Peraro:2016wsq}%
  \BibitemOpen
  \bibfield  {author} {\bibinfo {author} {\bibfnamefont {T.}~\bibnamefont {Peraro}},\ }\bibfield  {title} {\bibinfo {title} {{Scattering amplitudes over finite fields and multivariate functional reconstruction}},\ }\href {https://doi.org/10.1007/JHEP12(2016)030} {\bibfield  {journal} {\bibinfo  {journal} {JHEP}\ }\textbf {\bibinfo {volume} {12}},\ \bibinfo {pages} {030}},\ \Eprint {https://arxiv.org/abs/1608.01902} {arXiv:1608.01902 [hep-ph]} \BibitemShut {NoStop}%
\bibitem [{\citenamefont {Maierh{\"o}fer}\ \emph {et~al.}(2018)\citenamefont {Maierh{\"o}fer}, \citenamefont {Usovitsch},\ and\ \citenamefont {Uwer}}]{Maierhofer:2017gsa}%
  \BibitemOpen
  \bibfield  {author} {\bibinfo {author} {\bibfnamefont {P.}~\bibnamefont {Maierh{\"o}fer}}, \bibinfo {author} {\bibfnamefont {J.}~\bibnamefont {Usovitsch}},\ and\ \bibinfo {author} {\bibfnamefont {P.}~\bibnamefont {Uwer}},\ }\bibfield  {title} {\bibinfo {title} {{Kira{\textemdash}A Feynman integral reduction program}},\ }\href {https://doi.org/10.1016/j.cpc.2018.04.012} {\bibfield  {journal} {\bibinfo  {journal} {Comput. Phys. Commun.}\ }\textbf {\bibinfo {volume} {230}},\ \bibinfo {pages} {99} (\bibinfo {year} {2018})},\ \Eprint {https://arxiv.org/abs/1705.05610} {arXiv:1705.05610 [hep-ph]} \BibitemShut {NoStop}%
\bibitem [{\citenamefont {Maierh{\"o}fer}\ and\ \citenamefont {Usovitsch}(2018)}]{Maierhofer:2018gpa}%
  \BibitemOpen
  \bibfield  {author} {\bibinfo {author} {\bibfnamefont {P.}~\bibnamefont {Maierh{\"o}fer}}\ and\ \bibinfo {author} {\bibfnamefont {J.}~\bibnamefont {Usovitsch}},\ }\bibfield  {title} {\bibinfo {title} {{Kira 1.2 Release Notes}},\ }\href@noop {} {\  (\bibinfo {year} {2018})},\ \Eprint {https://arxiv.org/abs/1812.01491} {arXiv:1812.01491 [hep-ph]} \BibitemShut {NoStop}%
\bibitem [{\citenamefont {Maierh{\"o}fer}\ and\ \citenamefont {Usovitsch}(2020)}]{Maierhofer:2019goc}%
  \BibitemOpen
  \bibfield  {author} {\bibinfo {author} {\bibfnamefont {P.}~\bibnamefont {Maierh{\"o}fer}}\ and\ \bibinfo {author} {\bibfnamefont {J.}~\bibnamefont {Usovitsch}},\ }\bibfield  {title} {\bibinfo {title} {{Recent developments in Kira}},\ }\href {https://doi.org/10.23731/CYRM-2020-003.201} {\bibfield  {journal} {\bibinfo  {journal} {CERN Yellow Reports: Monographs}\ }\textbf {\bibinfo {volume} {3}},\ \bibinfo {pages} {201} (\bibinfo {year} {2020})}\BibitemShut {NoStop}%
\bibitem [{\citenamefont {Klappert}\ \emph {et~al.}(2021{\natexlab{a}})\citenamefont {Klappert}, \citenamefont {Lange}, \citenamefont {Maierh{\"o}fer},\ and\ \citenamefont {Usovitsch}}]{Klappert:2020nbg}%
  \BibitemOpen
  \bibfield  {author} {\bibinfo {author} {\bibfnamefont {J.}~\bibnamefont {Klappert}}, \bibinfo {author} {\bibfnamefont {F.}~\bibnamefont {Lange}}, \bibinfo {author} {\bibfnamefont {P.}~\bibnamefont {Maierh{\"o}fer}},\ and\ \bibinfo {author} {\bibfnamefont {J.}~\bibnamefont {Usovitsch}},\ }\bibfield  {title} {\bibinfo {title} {{Integral reduction with Kira 2.0 and finite field methods}},\ }\href {https://doi.org/10.1016/j.cpc.2021.108024} {\bibfield  {journal} {\bibinfo  {journal} {Comput. Phys. Commun.}\ }\textbf {\bibinfo {volume} {266}},\ \bibinfo {pages} {108024} (\bibinfo {year} {2021}{\natexlab{a}})},\ \Eprint {https://arxiv.org/abs/2008.06494} {arXiv:2008.06494 [hep-ph]} \BibitemShut {NoStop}%
\bibitem [{\citenamefont {Lange}\ \emph {et~al.}(2025)\citenamefont {Lange}, \citenamefont {Usovitsch},\ and\ \citenamefont {Wu}}]{Lange:2025fba}%
  \BibitemOpen
  \bibfield  {author} {\bibinfo {author} {\bibfnamefont {F.}~\bibnamefont {Lange}}, \bibinfo {author} {\bibfnamefont {J.}~\bibnamefont {Usovitsch}},\ and\ \bibinfo {author} {\bibfnamefont {Z.}~\bibnamefont {Wu}},\ }\bibfield  {title} {\bibinfo {title} {{Kira 3: integral reduction with efficient seeding and optimized equation selection}},\ }\href@noop {} {\  (\bibinfo {year} {2025})},\ \Eprint {https://arxiv.org/abs/2505.20197} {arXiv:2505.20197 [hep-ph]} \BibitemShut {NoStop}%
\bibitem [{\citenamefont {Klappert}\ and\ \citenamefont {Lange}(2020)}]{Klappert:2019emp}%
  \BibitemOpen
  \bibfield  {author} {\bibinfo {author} {\bibfnamefont {J.}~\bibnamefont {Klappert}}\ and\ \bibinfo {author} {\bibfnamefont {F.}~\bibnamefont {Lange}},\ }\bibfield  {title} {\bibinfo {title} {{Reconstructing rational functions with FireFly}},\ }\href {https://doi.org/10.1016/j.cpc.2019.106951} {\bibfield  {journal} {\bibinfo  {journal} {Comput. Phys. Commun.}\ }\textbf {\bibinfo {volume} {247}},\ \bibinfo {pages} {106951} (\bibinfo {year} {2020})},\ \Eprint {https://arxiv.org/abs/1904.00009} {arXiv:1904.00009 [cs.SC]} \BibitemShut {NoStop}%
\bibitem [{\citenamefont {Klappert}\ \emph {et~al.}(2021{\natexlab{b}})\citenamefont {Klappert}, \citenamefont {Klein},\ and\ \citenamefont {Lange}}]{Klappert:2020aqs}%
  \BibitemOpen
  \bibfield  {author} {\bibinfo {author} {\bibfnamefont {J.}~\bibnamefont {Klappert}}, \bibinfo {author} {\bibfnamefont {S.~Y.}\ \bibnamefont {Klein}},\ and\ \bibinfo {author} {\bibfnamefont {F.}~\bibnamefont {Lange}},\ }\bibfield  {title} {\bibinfo {title} {{Interpolation of dense and sparse rational functions and other improvements in FireFly}},\ }\href {https://doi.org/10.1016/j.cpc.2021.107968} {\bibfield  {journal} {\bibinfo  {journal} {Comput. Phys. Commun.}\ }\textbf {\bibinfo {volume} {264}},\ \bibinfo {pages} {107968} (\bibinfo {year} {2021}{\natexlab{b}})},\ \Eprint {https://arxiv.org/abs/2004.01463} {arXiv:2004.01463 [cs.MS]} \BibitemShut {NoStop}%
\bibitem [{\citenamefont {Magerya}(2022)}]{Magerya:2022hvj}%
  \BibitemOpen
  \bibfield  {author} {\bibinfo {author} {\bibfnamefont {V.}~\bibnamefont {Magerya}},\ }\bibfield  {title} {\bibinfo {title} {{Rational Tracer: a Tool for Faster Rational Function Reconstruction}},\ }\href@noop {} {\  (\bibinfo {year} {2022})},\ \Eprint {https://arxiv.org/abs/2211.03572} {arXiv:2211.03572 [physics.data-an]} \BibitemShut {NoStop}%
\bibitem [{\citenamefont {Guan}\ \emph {et~al.}(2020)\citenamefont {Guan}, \citenamefont {Liu},\ and\ \citenamefont {Ma}}]{Guan:2019bcx}%
  \BibitemOpen
  \bibfield  {author} {\bibinfo {author} {\bibfnamefont {X.}~\bibnamefont {Guan}}, \bibinfo {author} {\bibfnamefont {X.}~\bibnamefont {Liu}},\ and\ \bibinfo {author} {\bibfnamefont {Y.-Q.}\ \bibnamefont {Ma}},\ }\bibfield  {title} {\bibinfo {title} {{Complete reduction of integrals in two-loop five-light-parton scattering amplitudes}},\ }\href {https://doi.org/10.1088/1674-1137/44/9/093106} {\bibfield  {journal} {\bibinfo  {journal} {Chin. Phys. C}\ }\textbf {\bibinfo {volume} {44}},\ \bibinfo {pages} {093106} (\bibinfo {year} {2020})},\ \Eprint {https://arxiv.org/abs/1912.09294} {arXiv:1912.09294 [hep-ph]} \BibitemShut {NoStop}%
\bibitem [{\citenamefont {Liu}\ and\ \citenamefont {Ma}(2022)}]{Liu:2021wks}%
  \BibitemOpen
  \bibfield  {author} {\bibinfo {author} {\bibfnamefont {X.}~\bibnamefont {Liu}}\ and\ \bibinfo {author} {\bibfnamefont {Y.-Q.}\ \bibnamefont {Ma}},\ }\bibfield  {title} {\bibinfo {title} {{Multiloop corrections for collider processes using auxiliary mass flow}},\ }\href {https://doi.org/10.1103/PhysRevD.105.L051503} {\bibfield  {journal} {\bibinfo  {journal} {Phys. Rev. D}\ }\textbf {\bibinfo {volume} {105}},\ \bibinfo {pages} {L051503} (\bibinfo {year} {2022})},\ \Eprint {https://arxiv.org/abs/2107.01864} {arXiv:2107.01864 [hep-ph]} \BibitemShut {NoStop}%
\bibitem [{\citenamefont {Guan}\ \emph {et~al.}(2025)\citenamefont {Guan}, \citenamefont {Liu}, \citenamefont {Ma},\ and\ \citenamefont {Wu}}]{Guan:2024byi}%
  \BibitemOpen
  \bibfield  {author} {\bibinfo {author} {\bibfnamefont {X.}~\bibnamefont {Guan}}, \bibinfo {author} {\bibfnamefont {X.}~\bibnamefont {Liu}}, \bibinfo {author} {\bibfnamefont {Y.-Q.}\ \bibnamefont {Ma}},\ and\ \bibinfo {author} {\bibfnamefont {W.-H.}\ \bibnamefont {Wu}},\ }\bibfield  {title} {\bibinfo {title} {{Blade: A package for block-triangular form improved Feynman integrals decomposition}},\ }\href {https://doi.org/10.1016/j.cpc.2025.109538} {\bibfield  {journal} {\bibinfo  {journal} {Comput. Phys. Commun.}\ }\textbf {\bibinfo {volume} {310}},\ \bibinfo {pages} {109538} (\bibinfo {year} {2025})},\ \Eprint {https://arxiv.org/abs/2405.14621} {arXiv:2405.14621 [hep-ph]} \BibitemShut {NoStop}%
\bibitem [{\citenamefont {Wu}\ \emph {et~al.}(2024)\citenamefont {Wu}, \citenamefont {Boehm}, \citenamefont {Ma}, \citenamefont {Xu},\ and\ \citenamefont {Zhang}}]{Wu:2023upw}%
  \BibitemOpen
  \bibfield  {author} {\bibinfo {author} {\bibfnamefont {Z.}~\bibnamefont {Wu}}, \bibinfo {author} {\bibfnamefont {J.}~\bibnamefont {Boehm}}, \bibinfo {author} {\bibfnamefont {R.}~\bibnamefont {Ma}}, \bibinfo {author} {\bibfnamefont {H.}~\bibnamefont {Xu}},\ and\ \bibinfo {author} {\bibfnamefont {Y.}~\bibnamefont {Zhang}},\ }\bibfield  {title} {\bibinfo {title} {{NeatIBP 1.0, a package generating small-size integration-by-parts relations for Feynman integrals}},\ }\href {https://doi.org/10.1016/j.cpc.2023.108999} {\bibfield  {journal} {\bibinfo  {journal} {Comput. Phys. Commun.}\ }\textbf {\bibinfo {volume} {295}},\ \bibinfo {pages} {108999} (\bibinfo {year} {2024})},\ \Eprint {https://arxiv.org/abs/2305.08783} {arXiv:2305.08783 [hep-ph]} \BibitemShut {NoStop}%
\bibitem [{\citenamefont {Wu}\ \emph {et~al.}(2025)\citenamefont {Wu}, \citenamefont {B{\"o}hm}, \citenamefont {Ma}, \citenamefont {Usovitsch}, \citenamefont {Xu},\ and\ \citenamefont {Zhang}}]{Wu:2025aeg}%
  \BibitemOpen
  \bibfield  {author} {\bibinfo {author} {\bibfnamefont {Z.}~\bibnamefont {Wu}}, \bibinfo {author} {\bibfnamefont {J.}~\bibnamefont {B{\"o}hm}}, \bibinfo {author} {\bibfnamefont {R.}~\bibnamefont {Ma}}, \bibinfo {author} {\bibfnamefont {J.}~\bibnamefont {Usovitsch}}, \bibinfo {author} {\bibfnamefont {Y.}~\bibnamefont {Xu}},\ and\ \bibinfo {author} {\bibfnamefont {Y.}~\bibnamefont {Zhang}},\ }\bibfield  {title} {\bibinfo {title} {{Performing integration-by-parts reductions using NeatIBP 1.1 + Kira}},\ }\href@noop {} {\  (\bibinfo {year} {2025})},\ \Eprint {https://arxiv.org/abs/2502.20778} {arXiv:2502.20778 [hep-ph]} \BibitemShut {NoStop}%
\bibitem [{\citenamefont {von Manteuffel}\ and\ \citenamefont {Schabinger}(2015)}]{vonManteuffel:2014ixa}%
  \BibitemOpen
  \bibfield  {author} {\bibinfo {author} {\bibfnamefont {A.}~\bibnamefont {von Manteuffel}}\ and\ \bibinfo {author} {\bibfnamefont {R.~M.}\ \bibnamefont {Schabinger}},\ }\bibfield  {title} {\bibinfo {title} {{A novel approach to integration by parts reduction}},\ }\href {https://doi.org/10.1016/j.physletb.2015.03.029} {\bibfield  {journal} {\bibinfo  {journal} {Phys. Lett. B}\ }\textbf {\bibinfo {volume} {744}},\ \bibinfo {pages} {101} (\bibinfo {year} {2015})},\ \Eprint {https://arxiv.org/abs/1406.4513} {arXiv:1406.4513 [hep-ph]} \BibitemShut {NoStop}%
\bibitem [{\citenamefont {Gluza}\ \emph {et~al.}(2011)\citenamefont {Gluza}, \citenamefont {Kajda},\ and\ \citenamefont {Kosower}}]{Gluza:2010ws}%
  \BibitemOpen
  \bibfield  {author} {\bibinfo {author} {\bibfnamefont {J.}~\bibnamefont {Gluza}}, \bibinfo {author} {\bibfnamefont {K.}~\bibnamefont {Kajda}},\ and\ \bibinfo {author} {\bibfnamefont {D.~A.}\ \bibnamefont {Kosower}},\ }\bibfield  {title} {\bibinfo {title} {{Towards a Basis for Planar Two-Loop Integrals}},\ }\href {https://doi.org/10.1103/PhysRevD.83.045012} {\bibfield  {journal} {\bibinfo  {journal} {Phys. Rev. D}\ }\textbf {\bibinfo {volume} {83}},\ \bibinfo {pages} {045012} (\bibinfo {year} {2011})},\ \Eprint {https://arxiv.org/abs/1009.0472} {arXiv:1009.0472 [hep-th]} \BibitemShut {NoStop}%
\bibitem [{\citenamefont {Larsen}\ and\ \citenamefont {Zhang}(2016)}]{Larsen:2015ped}%
  \BibitemOpen
  \bibfield  {author} {\bibinfo {author} {\bibfnamefont {K.~J.}\ \bibnamefont {Larsen}}\ and\ \bibinfo {author} {\bibfnamefont {Y.}~\bibnamefont {Zhang}},\ }\bibfield  {title} {\bibinfo {title} {{Integration-by-parts reductions from unitarity cuts and algebraic geometry}},\ }\href {https://doi.org/10.1103/PhysRevD.93.041701} {\bibfield  {journal} {\bibinfo  {journal} {Phys. Rev. D}\ }\textbf {\bibinfo {volume} {93}},\ \bibinfo {pages} {041701} (\bibinfo {year} {2016})},\ \Eprint {https://arxiv.org/abs/1511.01071} {arXiv:1511.01071 [hep-th]} \BibitemShut {NoStop}%
\bibitem [{\citenamefont {Mastrolia}\ and\ \citenamefont {Mizera}(2019)}]{Mastrolia:2018uzb}%
  \BibitemOpen
  \bibfield  {author} {\bibinfo {author} {\bibfnamefont {P.}~\bibnamefont {Mastrolia}}\ and\ \bibinfo {author} {\bibfnamefont {S.}~\bibnamefont {Mizera}},\ }\bibfield  {title} {\bibinfo {title} {{Feynman Integrals and Intersection Theory}},\ }\href {https://doi.org/10.1007/JHEP02(2019)139} {\bibfield  {journal} {\bibinfo  {journal} {JHEP}\ }\textbf {\bibinfo {volume} {02}},\ \bibinfo {pages} {139}},\ \Eprint {https://arxiv.org/abs/1810.03818} {arXiv:1810.03818 [hep-th]} \BibitemShut {NoStop}%
\bibitem [{\citenamefont {Chestnov}\ \emph {et~al.}(2022)\citenamefont {Chestnov}, \citenamefont {Gasparotto}, \citenamefont {Mandal}, \citenamefont {Mastrolia}, \citenamefont {Matsubara-Heo}, \citenamefont {Munch},\ and\ \citenamefont {Takayama}}]{Chestnov:2022alh}%
  \BibitemOpen
  \bibfield  {author} {\bibinfo {author} {\bibfnamefont {V.}~\bibnamefont {Chestnov}}, \bibinfo {author} {\bibfnamefont {F.}~\bibnamefont {Gasparotto}}, \bibinfo {author} {\bibfnamefont {M.~K.}\ \bibnamefont {Mandal}}, \bibinfo {author} {\bibfnamefont {P.}~\bibnamefont {Mastrolia}}, \bibinfo {author} {\bibfnamefont {S.~J.}\ \bibnamefont {Matsubara-Heo}}, \bibinfo {author} {\bibfnamefont {H.~J.}\ \bibnamefont {Munch}},\ and\ \bibinfo {author} {\bibfnamefont {N.}~\bibnamefont {Takayama}},\ }\bibfield  {title} {\bibinfo {title} {{Macaulay matrix for Feynman integrals: linear relations and intersection numbers}},\ }\href {https://doi.org/10.1007/JHEP09(2022)187} {\bibfield  {journal} {\bibinfo  {journal} {JHEP}\ }\textbf {\bibinfo {volume} {09}},\ \bibinfo {pages} {187}},\ \Eprint {https://arxiv.org/abs/2204.12983} {arXiv:2204.12983 [hep-th]} \BibitemShut {NoStop}%
\bibitem [{\citenamefont {Brunello}\ \emph {et~al.}(2025)\citenamefont {Brunello}, \citenamefont {Chestnov},\ and\ \citenamefont {Mastrolia}}]{Brunello:2024tqf}%
  \BibitemOpen
  \bibfield  {author} {\bibinfo {author} {\bibfnamefont {G.}~\bibnamefont {Brunello}}, \bibinfo {author} {\bibfnamefont {V.}~\bibnamefont {Chestnov}},\ and\ \bibinfo {author} {\bibfnamefont {P.}~\bibnamefont {Mastrolia}},\ }\bibfield  {title} {\bibinfo {title} {{Intersection numbers from companion tensor algebra}},\ }\href {https://doi.org/10.1007/JHEP07(2025)045} {\bibfield  {journal} {\bibinfo  {journal} {JHEP}\ }\textbf {\bibinfo {volume} {07}},\ \bibinfo {pages} {045}},\ \Eprint {https://arxiv.org/abs/2408.16668} {arXiv:2408.16668 [hep-th]} \BibitemShut {NoStop}%
\bibitem [{\citenamefont {Bern}\ \emph {et~al.}(2025)\citenamefont {Bern}, \citenamefont {Herrmann}, \citenamefont {Roiban}, \citenamefont {Ruf}, \citenamefont {Smirnov}, \citenamefont {Smith},\ and\ \citenamefont {Zeng}}]{Bern:2025wyd}%
  \BibitemOpen
  \bibfield  {author} {\bibinfo {author} {\bibfnamefont {Z.}~\bibnamefont {Bern}}, \bibinfo {author} {\bibfnamefont {E.}~\bibnamefont {Herrmann}}, \bibinfo {author} {\bibfnamefont {R.}~\bibnamefont {Roiban}}, \bibinfo {author} {\bibfnamefont {M.~S.}\ \bibnamefont {Ruf}}, \bibinfo {author} {\bibfnamefont {A.~V.}\ \bibnamefont {Smirnov}}, \bibinfo {author} {\bibfnamefont {S.}~\bibnamefont {Smith}},\ and\ \bibinfo {author} {\bibfnamefont {M.}~\bibnamefont {Zeng}},\ }\bibfield  {title} {\bibinfo {title} {{Scattering Amplitudes and Conservative Binary Dynamics at $O(G^5)$ without Self-Force Truncation}},\ }\href@noop {} {\  (\bibinfo {year} {2025})},\ \Eprint {https://arxiv.org/abs/2512.23654} {arXiv:2512.23654 [hep-th]} \BibitemShut {NoStop}%
\bibitem [{\citenamefont {Driesse}\ \emph {et~al.}(2026)\citenamefont {Driesse}, \citenamefont {Jakobsen}, \citenamefont {Mogull}, \citenamefont {Nega}, \citenamefont {Plefka}, \citenamefont {Sauer},\ and\ \citenamefont {Usovitsch}}]{Driesse:2026qiz}%
  \BibitemOpen
  \bibfield  {author} {\bibinfo {author} {\bibfnamefont {M.}~\bibnamefont {Driesse}}, \bibinfo {author} {\bibfnamefont {G.~U.}\ \bibnamefont {Jakobsen}}, \bibinfo {author} {\bibfnamefont {G.}~\bibnamefont {Mogull}}, \bibinfo {author} {\bibfnamefont {C.}~\bibnamefont {Nega}}, \bibinfo {author} {\bibfnamefont {J.}~\bibnamefont {Plefka}}, \bibinfo {author} {\bibfnamefont {B.}~\bibnamefont {Sauer}},\ and\ \bibinfo {author} {\bibfnamefont {J.}~\bibnamefont {Usovitsch}},\ }\bibfield  {title} {\bibinfo {title} {{Conservative Black Hole Scattering at Fifth Post-Minkowskian and Second Self-Force Order}},\ }\href@noop {} {\  (\bibinfo {year} {2026})},\ \Eprint {https://arxiv.org/abs/2601.16256} {arXiv:2601.16256 [hep-th]} \BibitemShut {NoStop}%
\bibitem [{\citenamefont {Liu}\ \emph {et~al.}(2025)\citenamefont {Liu}, \citenamefont {Matija{\v{s}}i{\'c}}, \citenamefont {Miczajka}, \citenamefont {Xu}, \citenamefont {Xu},\ and\ \citenamefont {Zhang}}]{Liu:2024ont}%
  \BibitemOpen
  \bibfield  {author} {\bibinfo {author} {\bibfnamefont {Y.}~\bibnamefont {Liu}}, \bibinfo {author} {\bibfnamefont {A.}~\bibnamefont {Matija{\v{s}}i{\'c}}}, \bibinfo {author} {\bibfnamefont {J.}~\bibnamefont {Miczajka}}, \bibinfo {author} {\bibfnamefont {Y.}~\bibnamefont {Xu}}, \bibinfo {author} {\bibfnamefont {Y.}~\bibnamefont {Xu}},\ and\ \bibinfo {author} {\bibfnamefont {Y.}~\bibnamefont {Zhang}},\ }\bibfield  {title} {\bibinfo {title} {{Analytic computation of three-loop five-point Feynman integrals}},\ }\href {https://doi.org/10.1103/qrk2-cym5} {\bibfield  {journal} {\bibinfo  {journal} {Phys. Rev. D}\ }\textbf {\bibinfo {volume} {112}},\ \bibinfo {pages} {016021} (\bibinfo {year} {2025})},\ \Eprint {https://arxiv.org/abs/2411.18697} {arXiv:2411.18697 [hep-ph]} \BibitemShut {NoStop}%
\bibitem [{\citenamefont {Chicherin}\ \emph {et~al.}(2025)\citenamefont {Chicherin}, \citenamefont {Wu}, \citenamefont {Wu}, \citenamefont {Xu}, \citenamefont {Zhang},\ and\ \citenamefont {Zhang}}]{Chicherin:2025mvc}%
  \BibitemOpen
  \bibfield  {author} {\bibinfo {author} {\bibfnamefont {D.}~\bibnamefont {Chicherin}}, \bibinfo {author} {\bibfnamefont {Y.}~\bibnamefont {Wu}}, \bibinfo {author} {\bibfnamefont {Z.}~\bibnamefont {Wu}}, \bibinfo {author} {\bibfnamefont {Y.}~\bibnamefont {Xu}}, \bibinfo {author} {\bibfnamefont {S.-Q.}\ \bibnamefont {Zhang}},\ and\ \bibinfo {author} {\bibfnamefont {Y.}~\bibnamefont {Zhang}},\ }\bibfield  {title} {\bibinfo {title} {{Complete computation of all three-loop five-point massless planar integrals}},\ }\href@noop {} {\  (\bibinfo {year} {2025})},\ \Eprint {https://arxiv.org/abs/2512.17330} {arXiv:2512.17330 [hep-ph]} \BibitemShut {NoStop}%
\bibitem [{\citenamefont {Smirnov}\ and\ \citenamefont {Smirnov}(2006{\natexlab{a}})}]{Smirnov:2005ky}%
  \BibitemOpen
  \bibfield  {author} {\bibinfo {author} {\bibfnamefont {A.~V.}\ \bibnamefont {Smirnov}}\ and\ \bibinfo {author} {\bibfnamefont {V.~A.}\ \bibnamefont {Smirnov}},\ }\bibfield  {title} {\bibinfo {title} {{Applying Grobner bases to solve reduction problems for Feynman integrals}},\ }\href {https://doi.org/10.1088/1126-6708/2006/01/001} {\bibfield  {journal} {\bibinfo  {journal} {JHEP}\ }\textbf {\bibinfo {volume} {01}},\ \bibinfo {pages} {001}},\ \Eprint {https://arxiv.org/abs/hep-lat/0509187} {arXiv:hep-lat/0509187} \BibitemShut {NoStop}%
\bibitem [{\citenamefont {Tarasov}(1998)}]{Tarasov:1998nx}%
  \BibitemOpen
  \bibfield  {author} {\bibinfo {author} {\bibfnamefont {O.~V.}\ \bibnamefont {Tarasov}},\ }\bibfield  {title} {\bibinfo {title} {{Reduction of Feynman graph amplitudes to a minimal set of basic integrals}},\ }\href@noop {} {\bibfield  {journal} {\bibinfo  {journal} {Acta Phys. Polon. B}\ }\textbf {\bibinfo {volume} {29}},\ \bibinfo {pages} {2655} (\bibinfo {year} {1998})},\ \Eprint {https://arxiv.org/abs/hep-ph/9812250} {arXiv:hep-ph/9812250} \BibitemShut {NoStop}%
\bibitem [{\citenamefont {Gerdt}(2004)}]{Gerdt:2004kt}%
  \BibitemOpen
  \bibfield  {author} {\bibinfo {author} {\bibfnamefont {V.~P.}\ \bibnamefont {Gerdt}},\ }\bibfield  {title} {\bibinfo {title} {{Grobner bases in perturbative calculations}},\ }\href {https://doi.org/10.1016/j.nuclphysbps.2004.09.011} {\bibfield  {journal} {\bibinfo  {journal} {Nucl. Phys. B Proc. Suppl.}\ }\textbf {\bibinfo {volume} {135}},\ \bibinfo {pages} {232} (\bibinfo {year} {2004})},\ \Eprint {https://arxiv.org/abs/hep-ph/0501053} {arXiv:hep-ph/0501053} \BibitemShut {NoStop}%
\bibitem [{\citenamefont {Smirnov}\ and\ \citenamefont {Smirnov}(2006{\natexlab{b}})}]{Smirnov:2006wh}%
  \BibitemOpen
  \bibfield  {author} {\bibinfo {author} {\bibfnamefont {A.~V.}\ \bibnamefont {Smirnov}}\ and\ \bibinfo {author} {\bibfnamefont {V.~A.}\ \bibnamefont {Smirnov}},\ }\bibfield  {title} {\bibinfo {title} {{S-bases as a tool to solve reduction problems for Feynman integrals}},\ }\href {https://doi.org/10.1016/j.nuclphysbps.2006.09.032} {\bibfield  {journal} {\bibinfo  {journal} {Nucl. Phys. B Proc. Suppl.}\ }\textbf {\bibinfo {volume} {160}},\ \bibinfo {pages} {80} (\bibinfo {year} {2006}{\natexlab{b}})},\ \Eprint {https://arxiv.org/abs/hep-ph/0606247} {arXiv:hep-ph/0606247} \BibitemShut {NoStop}%
\bibitem [{\citenamefont {Smirnov}(2006)}]{Smirnov:2006tz}%
  \BibitemOpen
  \bibfield  {author} {\bibinfo {author} {\bibfnamefont {A.~V.}\ \bibnamefont {Smirnov}},\ }\bibfield  {title} {\bibinfo {title} {{An Algorithm to construct Grobner bases for solving integration by parts relations}},\ }\href {https://doi.org/10.1088/1126-6708/2006/04/026} {\bibfield  {journal} {\bibinfo  {journal} {JHEP}\ }\textbf {\bibinfo {volume} {04}},\ \bibinfo {pages} {026}},\ \Eprint {https://arxiv.org/abs/hep-ph/0602078} {arXiv:hep-ph/0602078} \BibitemShut {NoStop}%
\bibitem [{\citenamefont {Feng}(2023)}]{Feng:2022hyg}%
  \BibitemOpen
  \bibfield  {author} {\bibinfo {author} {\bibfnamefont {B.}~\bibnamefont {Feng}},\ }\bibfield  {title} {\bibinfo {title} {{Generation function for one-loop tensor reduction}},\ }\href {https://doi.org/10.1088/1572-9494/aca253} {\bibfield  {journal} {\bibinfo  {journal} {Commun. Theor. Phys.}\ }\textbf {\bibinfo {volume} {75}},\ \bibinfo {pages} {025203} (\bibinfo {year} {2023})},\ \Eprint {https://arxiv.org/abs/2209.09517} {arXiv:2209.09517 [hep-ph]} \BibitemShut {NoStop}%
\bibitem [{\citenamefont {Guan}\ \emph {et~al.}(2023)\citenamefont {Guan}, \citenamefont {Li},\ and\ \citenamefont {Ma}}]{Guan:2023avw}%
  \BibitemOpen
  \bibfield  {author} {\bibinfo {author} {\bibfnamefont {X.}~\bibnamefont {Guan}}, \bibinfo {author} {\bibfnamefont {X.}~\bibnamefont {Li}},\ and\ \bibinfo {author} {\bibfnamefont {Y.-Q.}\ \bibnamefont {Ma}},\ }\bibfield  {title} {\bibinfo {title} {{Exploring the linear space of Feynman integrals via generating functions}},\ }\href {https://doi.org/10.1103/PhysRevD.108.034027} {\bibfield  {journal} {\bibinfo  {journal} {Phys. Rev. D}\ }\textbf {\bibinfo {volume} {108}},\ \bibinfo {pages} {034027} (\bibinfo {year} {2023})},\ \Eprint {https://arxiv.org/abs/2306.02927} {arXiv:2306.02927 [hep-ph]} \BibitemShut {NoStop}%
\bibitem [{\citenamefont {Smith}\ and\ \citenamefont {Zeng}(2025)}]{Smith:2025xes}%
  \BibitemOpen
  \bibfield  {author} {\bibinfo {author} {\bibfnamefont {S.}~\bibnamefont {Smith}}\ and\ \bibinfo {author} {\bibfnamefont {M.}~\bibnamefont {Zeng}},\ }\bibfield  {title} {\bibinfo {title} {{Feynman Integral Reduction using Syzygy-Constrained Symbolic Reduction Rules}},\ }\href@noop {} {\  (\bibinfo {year} {2025})},\ \Eprint {https://arxiv.org/abs/2507.11140} {arXiv:2507.11140 [hep-th]} \BibitemShut {NoStop}%
\bibitem [{Note1()}]{Note1}%
  \BibitemOpen
  \bibinfo {note} {$\protect \vec {a}\geq 0$, means that every component of the vector is non-negative integer and $\protect \vec {a}\not \geq 0$ means that some components of the vector will be negative integer.}\BibitemShut {Stop}%
\bibitem [{Note2()}]{Note2}%
  \BibitemOpen
  \bibinfo {note} {Later, a zero-index operator could also mean a linear combination of zero-index operators}\BibitemShut {NoStop}%
\end{thebibliography}%
\newpage

%%%%%%%%%%%%%%%%%%%%%
\appendix

\begin{center}
\textbf{Supplemental Material}
\end{center}

In the supplemental material, we provide  details of the examples in this paper, as well as the two-loop four-point planar integral reduction example.

%%%%%%%%%%%%%%%%%
\section{The topsector of sunset}
%%%%%%%%%%%%%%%

For the sunset, the propagators and ISPs are 
\bea \mathcal{D}_1 &= &(\ell_1^2-m_1^2),~~~ \mathcal{D}_2=(\ell_2^2-m_2^2),\nn \mathcal{D}_3 &= &((\ell_1+\ell_2-K)^2-m_3^2),\nn
\mathcal{D}_4&= &\ell_1\cdot K,~~~~ \mathcal{D}_5=\ell_2\cdot K.~~~\label{sunset-1-2}\eea
Among  $2^3=8$ possible generating function, only following four sectors are nonzero 
\bea G_{111}, G_{110}, G_{101}, G_{011}~~~\label{sunset-1-3}\eea
where for simplicity $\mu_4=\mu_5=0$ have been implied. Among  $2^3=8$ possible generating function, only following four sectors are nonzero 
\bea G_{111}, G_{110}, G_{101}, G_{011}~~~\label{sunset-1-3}\eea
where for simplicity $\mu_4=\mu_5=0$ have been implied.

{\bf The first round of computation:} The initial seed for top-sector $G_{11100}=G_{111}$ is six degree-2 DE's from 
fundamental IBP relations. Focus on the degree-2 operators in these equations, we have following coefficient matrix 
{
	\bea \left( \begin{array}{c| c |  c |c |c |c |l} [1,4] &  [2,4]&[3,4] & [1,5] &[2,5] &[3,5] & {\rm  IBP}\\  \hline
		&  &  &  &  & 2 & {d\over \d\ell_1}\ell_1  \\  \hline 
		-2 &  & -2  & -2 &  &   & {d\over \d\ell_1}\ell_2  \\  \hline 
		-2 &  & -2 &  &  & -2  & {d\over \d\ell_1}K  \\  \hline 
		& -2 &  &  & -2 & -2  & {d\over \d\ell_1}\ell_1   \\  \hline 
		&  & 2 &  &  &  & {d\over \d\ell_2}\ell_2   \\   \hline 
		&  & -2 &  & -2 & -2 & {d\over \d\ell_2}K 
	\end{array}\right)~~~~~~~\label{sunset-B1-1-1}\eea}
where $[a,b]\equiv {\d\over \d\eta_a} {\d\over \d\eta_b}$.

Doing the Gauss elimination, we can get the proper combinations of these six equations. After  solving them, we get six T1A-type reduction rules 
for  six degree-2 operators ${\d\over \d \eta_i} {\d\over \d \eta_j}$,
$i=4,5; j=1,2,3$, which are (where $f_{\pm \pm \pm}=\pm m_1^2\pm m_2^2\pm m_3^2-K^2$)
% (see \eqref{sunset-2-1-4} to \eqref{sunset-2-6-6} for their explicit expressions). For example, the IBP relation $0= \int\prod_{i=1}^{L}\frac{\md^{D}\ell_i}{\mi\pi^{D/2}} {\d\over \d\ell_1}\cdot \left\{ \ell_1 e^{\sum_{j=4,5} \eta_j s_0^{-1}\mathcal{D}_j}\prod_{i=1}^3 {1\over \mathcal{D}_i-\eta_i s_0 }\right\}$ produces
%
\bea  &  & \left\{  2 {\d\over \d \eta_5} {\d\over \d \eta_3} G_{111} \right\}   =    \left\{ {2m_1^2 \over s_0} {\d \over \d \eta_1} G_{111}+{f_{+-+}\over s_0}{\d\over \d \eta_3} G_{111}  \right\}\nn
& &    -     \left\{  (D-3) G_{111}+ \eta_4{\d\over \d \eta_4} G_{111}-2 \eta_1 {\d \over \d \eta_1} G_{111}\right.\nn & & \left. -(\eta_3+\eta_1-\eta_2){\d\over \d \eta_3} G_{111}   \right\}  \nn
& &  -     \left\{ -{1\over s_0} {\d\over \d \eta_3} G_{011}|_{\eta_1=0}+{1\over s_0}{\d\over \d \eta_3} G_{101}|_{\eta_2=0}  \right\}, ~~~\label{sunset-2-1-4}   \eea
%
%where $f_{\pm \pm \pm}=\pm m_1^2\pm m_2^2\pm m_3^2-K^2$.
%Using six DEs coming from six fundamental IBP relations, we can  solve  six degree two operators as  
%
\bea & &  2 {\d\over \d \eta_5} {\d\over \d \eta_3} G_{111}= -\left\{   {2m_1^2 \over s_0} {\d \over \d \eta_1} G_{111}+{f_{+-+}\over s_0}{\d\over \d \eta_3} G_{111} \right\}\nn & &  + \left\{ (D-3) G_{111} + \eta_4{\d\over \d \eta_4} G_{111}-2 \eta_1 {\d \over \d \eta_1} G_{111}\right.\nn & & \left.-(\eta_3+\eta_1-\eta_2){\d\over \d \eta_3} G_{111}  \right\} ~~~\label{sunset-2-1-4}\eea
\bea  & &  2 {\d \over \d \eta_5}{\d \over \d \eta_1} G_{111}=  \left\{  -{ f_{+-+}\over s_0}{\d \over \d \eta_1} G_{111} -{2m_3^2\over s_0}{\d\over \d \eta_3} G_{111}  \right\}\nn & &  +\left\{  (D-3) G_{111} + \eta_4{\d\over \d \eta_4} G_{111}+ \eta_4{\d\over \d \eta_5}G_{111}\right.\nn & & \left.+(-\eta_3-\eta_1+\eta_2) {\d \over \d \eta_1} G_{111} -2 \eta_3{\d\over \d \eta_3} G_{111} \right\}\nn & &  + \left\{ -\eta_4 K^2 s_0^{-1}G_{111} \right\}~~~~\label{sunset-2-2-6} 
\eea
\bea  & & 2 {\d \over \d \eta_4}{\d \over \d \eta_1}  G_{111}= \left\{ {2m_3^2\over s_0}{\d \over \d \eta_3} G_{111}+ {2m_1^2 \over s_0} {\d \over \d \eta_1} G_{111}\right.\nn & & \left.+{2m_2^2\over s_0}{\d\over \d \eta_2} G_{111}\right\} +\left\{ -2(D-3) G_{111}- \eta_4{\d\over \d \eta_4} G_{111}\right.\nn & & \left.-\eta_5{\d\over \d \eta_5} G_{111} +2 \eta_1 {\d \over \d \eta_1} G_{111}+ 2 \eta_2 {\d \over \d \eta_2} G_{111}\right.\nn & & \left.+2\eta_3{\d\over \d \eta_3} G_{111}\right\}+  \left\{\eta_4 K^2 s_0^{-1}G_{111}\right\}~~~~\label{sunset-2-3-6}
\eea
\bea & & 2 {\d \over \d \eta_4}{\d \over \d \eta_2} G_{111}=\left\{-{ f_{-++}\over s_0}{\d \over \d \eta_2} G_{111} -{2 m_3^2\over s_0}{\d\over \d \eta_3} G_{111}\right\}\nn & &  + \left\{(D-3) G_{111} + \eta_5{\d\over \d \eta_4} G_{111}+ \eta_5{\d\over \d \eta_5}G_{111} \right.\nn & & \left.+(-\eta_3+\eta_1-\eta_2) {\d \over \d \eta_2} G_{111}-2 \eta_3{\d\over \d \eta_3} G_{111}\right\} \nn & & +\left\{ -\eta_5 K^2 s_0^{-1}G_{111} \right\}~~~~\label{sunset-2-4-6}
\eea
\bea & & 2 {\d\over \d \eta_4} {\d\over \d \eta_3} G_{111}= -\left\{  {2m_2^2\over  s_0} {\d \over \d \eta_2} G_{111}+{f_{-++}\over s_0}{\d\over \d \eta_3} G_{111} \right\} \nn & &  +\left\{  (D-3) G_{111} + \eta_5{\d\over \d \eta_5} G_{111}-2 \eta_2 {\d \over \d \eta_2} G_{111}\right.\nn & & \left.-(\eta_3-\eta_1+\eta_2){\d\over \d \eta_3} G_{111}  \right\} ~~~\label{sunset-2-5-4}\eea
\bea  & & 2 {\d \over \d \eta_5}{\d \over \d \eta_2} G_{111}= \left\{ {2m_3^2\over s_0}{\d \over \d \eta_3} G_{111}+ {2m_1^2 \over s_0} {\d \over \d \eta_1} G_{111}\right.\nn & & \left.+{2m_2^2\over s_0}{\d\over \d \eta_2} G_{111} \right\}\ +\left\{ -2(D-3) G_{111}- \eta_4{\d\over \d \eta_4} G_{111}\right.\nn & & \left.-\eta_5{\d\over \d \eta_5} G_{111} +2 \eta_1 {\d \over \d \eta_1} G_{111}+ 2 \eta_2 {\d \over \d \eta_2} G_{111}\right.\nn & & \left.+2\eta_3{\d\over \d \eta_3} G_{111} \right\} + \left\{  \eta_5 K^2 s_0^{-1}G_{111} \right\}~~~~\label{sunset-2-6-6}
\eea
where since we focus on the top sector only, contributions from subsectors have been neglected. Also, we have written into the form with degree separated to demonstrate our idea. 

Using them, we can reduce any point except the following two sets
${\cal U}_1=(n_1,n_2,n_3,0,0)$ and ${\cal U}_2=(0,0,0,n_4,n_5)$. Since there are still infinite number of irreducible points, we need go to the next round of computation. 

{\bf The second round of computation:} 
Acting on ${\d\over \d\eta_i}$ on above six equations, we will get $30$ degree-3 equations. However,
among them, only $9$ provides nontrivial new equations. To see it, let us consider the equation produced by ${\d\over \d\eta_4}$ on the reduction rule ${\d\over \d\eta_5}{\d\over \d\eta_1}=...$. We get a degree-3 operator ${\d\over \d\eta_4}{\d\over \d\eta_5}{\d\over \d\eta_1}$, which is
the descendant of another operator ${\d\over \d\eta_4}{\d\over \d\eta_1}$ and can be simplified using the reduction rule \eref{sunset-2-3-6} with the action ${\d\over \d\eta_5}$. After done it, the equation becomes degree-2 with following operators ${\d\over \d\eta_4}{\d\over \d\eta_3}$, ${\d\over \d\eta_4}{\d\over \d\eta_1}$, ${\d\over \d\eta_5}{\d\over \d\eta_1}$, ${\d\over \d\eta_5}{\d\over \d\eta_2}$ and ${\d\over \d\eta_5}{\d\over \d\eta_3}$. All of them can be simplified and we reach a new degree-1 equation
% by using descendant reduction rules of above six reduction rules, one can check that we get $9$ degree $2$ equations, where one of them is
%
\bea 
0& = &\left\{   -\eta_4 {\d^2 \over \d \eta_4^2}G_{111}-\eta_5 {\d^2 \over \d \eta_5^2}G_{111} -2\eta_4 {\d\over \d \eta_4}{\d \over \d \eta_5} G_{111}\right.\nn & & \left. + {2m_1^2(m_2^2-m_3^2)\over s_0^2}{\d \over \d \eta_1}G_{111} +{m_2^2 f_{++-}\over s_0^2}{\d \over \d \eta_2}G_{111}
\right.\nn & & \left. -{m_3^2 f_{+-+}\over s_0^2}{\d \over \d \eta_3}G_{111}-(D-2){\d \over \d \eta_4}G_{111}\right.\nn & & \left.+(4-2D) {\d \over \d \eta_5}G_{111}    \right\} +...
%  
%& &+ \left\{    + {2(m_2^2-m_3^2)\eta_1+2 m_1^2(\eta_2-\eta_3)\over  s_0}  {\d \over \d \eta_1}G_{111}\right.\nn & & \left. - {m_2^2(\eta_1+\eta_2-\eta_3)+ f_{++-}\eta_2\over  s_0}{\d \over \d \eta_2}G_{111}\right. \nn
%
%& & + {-f_{+-+}\eta_3-m_3^2 (\eta_1-\eta_2+\eta_3)\over s_0}{\d \over \d \eta_3}G_{111}\nn & &  + {2K^2-f_{-+-}\over 2 s_0}\eta_4{\d \over \d \eta_4}G_{111}\nn
%
%& & \left.  +{2(K^2+m_1^2)\eta_4- f_{++-}\eta_5\over 2 s_0} {\d \over \d \eta_5}G_{111}\right.\nn & & \left. +{(D-2)K^2-(D-3)(m_2^2-m_3^2)\over s_0} G_{111}   \right\} \nn
%
%  
% the degree -1
%& &+ \left\{  + 2(\eta_2-\eta_3)\eta_1 {\d \over \d \eta_1}G_{111}+ (\eta_1+\eta_2-\eta_3)\eta_2{\d \over \d \eta_2}G_{111}\right.\nn & & \left. -(\eta_1-\eta_2+\eta_3)\eta_3{\d \over \d \eta_3}G_{111}+{(\eta_1-\eta_2+\eta_3)\over 2}\eta_4{\d \over \d \eta_4}G_{111}\right.\nn & & + (\eta_1\eta_4-{1\over 2} \eta_5(\eta_1+\eta_2-\eta_3))  {\d \over \d \eta_5}G_{111}\nn
%
%& & \left. +({ K^2 f_{--+}\over 2 s_0^2}\eta_4+{K^2 m_2^2\over s_0^2}\eta_5-(D-3) (\eta_2-\eta_3))G_{111}   \right\} \nn
%
%
% the degree -2
%
%& &+ \left\{  +({K^2\over s_0}\eta_2\eta_5-{K^2\over 2 s_0} \eta_4(\eta_1+\eta_2-\eta_3)) G_{111}   \right\} 
~~~\label{sunset-I145-1}\eea
where we have kept only the leading degree part. 

As can see from this example, only after this simplification, the dust coming from the known reduction rules will be cleaned away and new unsolved operators will appear manifestly as the leading degree operators. Solving these new appeared operators we will get new reduction rules. Doing it round by round, eventually we will 
find the complete reduction rules.

%the simplification is crucial step. Only after the simplification, As we will show in various examples, above simplification is the one of most crucial parts of the whole algorithm. Since only after this simplification, the dust coming from the known reduction rules will be cleaned away and new unsolved operators will appear manifestly. Solving these new appeared operators we will get new reduction rules. Doing it round by round, eventually we will  find the complete reduction rules.

These equations are, in fact, the consequence of integrability conditions $[{\d\over d\eta_i}, {\d\over \d\eta_j}]=0$. Above  $9$ new equations can be divided into three groups. The first group is given by ${\d\over \d\eta_4} \left({\d\over \d\eta_5}{\d\over \d\eta_i} \right)-{\d\over \d\eta_5} \left({\d\over \d\eta_4}{\d\over \d\eta_i} \right)=0$ with $i=1,2,3$. We will denote the corresponding equations as (I-i) (\eqref{sunset-I145-1} is (I-1)).  The second group is given by  ${\d\over \d\eta_j} \left({\d\over \d\eta_4}{\d\over \d\eta_i} \right)-{\d\over \d\eta_i} \left({\d\over \d\eta_4}{\d\over \d\eta_j} \right)=0$ with $1\leq i<j\leq 3$, which will be denoted as (II-ij).  The third group is given by  ${\d\over \d\eta_j} \left({\d\over \d\eta_5}{\d\over \d\eta_i} \right)-{\d\over \d\eta_i} \left({\d\over \d\eta_5}{\d\over \d\eta_j} \right)=0$ with $1\leq i<j\leq 3$, which will be denoted as (III-ij).  These $9$ equations are not independent since ${\text{ (II-12)}}+{\text{ (III-12)}}-{\text{ (II-23)}}-{\text{ (III-13)}}=0$.

Another important feature is that the degree of equations depend on the masses. For example, 
the (II-23) is 
\bea 
%% degree two part
& & {2 m_2^2 \over s_0 }{\d^2 \over \d \eta_2^2} G_{111}-{2 m_3^2 \over s_0 }{\d^2 \over \d \eta_3^2} G_{111}\nn
%
%%% degree one part 
& = &  \left\{   -(D-4) {\d\over \d \eta_3} G_{111}+ (D-4){\d\over \d \eta_2} G_{111}+2\eta_3 {\d^2 \over \d \eta_3^2} G_{111}\right.\nn & & \left. -2\eta_2 {\d^2 \over \d \eta_2^2} G_{111} \right\} +... ~~~\label{sunset-I234-1}
\eea
which will be degree-1 for massless case. 
 This difference will lead to different master integrals. 
 
{\bf Solving module of second round for massless case:} For $m_i=0, i=1,2,3$, we have $3$ degree-2 equations with coefficient matrix as 
\bea \left( \begin{array}{c| c |c |l} [1,2] &  [1,3] & [2,3] &  \\  \hline
	{K^2\over s_0} &   &   &  {\rm (II-12)}\\  \hline 
	& {K^2\over s_0}    &   & {\rm (II-13)}\\  \hline 
	%
%	-{K^2\over s_0} &   &   &  I_{\eref{sunset-I125-simpler}}\\  \hline
	%
	& & -{K^2\over s_0}    &  {\rm (III-23)}
\end{array}\right)~~~~~\label{sunset-massless-B-1}
\eea
and $5$ degree-1 equations with coefficient matrix
is 
{\small
\bea \left( \begin{array}{c| c |c |c| c| l} {\d\over \d\eta_1} & {\d\over \d\eta_2} &   {\d\over \d\eta_3} & {\d\over \d\eta_4} & {\d\over \d\eta_5} &  \\  \hline
	&   &   & -\WH O_{0;4,1}  &  \WH O_{0;5,2}  &  {\rm (I-1)} \\  \hline
	&   &   & \WH O_{0;4,2}  &  -\WH O_{0;5,1}   &  {\rm (I-2)} \\  \hline
	&   &   &  -\WH O_{0;4,1} &  \WH O_{0;5,1}  &  {\rm (I-3)}\\  \hline
	& \WH O_{0;2,1}  & -\WH O_{0;3,1}  &   &    &  {\rm (II-23)} \\  \hline
	-\WH O_{0;1,1} &   & \WH O_{0;3,1}  &   &    &  {\rm (III-13)} \\  %\hline
	%%
%	-\WH O_{0;1,1} & \WH O_{0;2,1}  &   &   &    &  I_{\eref{sunset-I124plus125-m=0}} \\  
	%
\end{array}\right)~~~~~\label{sunset-massless-B-3}
\eea }
with 
\bea \WH O_{0;a,1} & = & (D-2)+ \eta_a {\d\over \d\eta_a},~~,a=4,5\nn
\WH O_{0;4,2}&  = &  4-2D-\eta_4 {\d\over \d\eta_4}-2 \eta_5 {\d\over \d\eta_5}  \nn
%
%\WH O_{0;5,1} & = & (D-2)+ \eta_5 {\d\over \d\eta_5},\nn
%
\WH O_{0;5,2} & = &  4-2D-\eta_5 {\d\over \d\eta_5}-2 \eta_4 {\d\over \d\eta_4}\nn
\WH O_{0;i,1} & = & (D-4)- 2\eta_i {\d\over \d\eta_i},~i=1,2,3~~~~~\label{sunset-massless-B-3-O}\eea
Doing Gauss elimination for the second and third row gives 
{\small
\bea \left( \begin{array}{c| c |c |c| c| l} {\d\over \d\eta_1} & {\d\over \d\eta_2} &   {\d\over \d\eta_3} & {\d\over \d\eta_4} & {\d\over \d\eta_5} &  \\  \hline
	%
%	&   &   & -\WH O_{0;4,1}  &  \WH O_{0;5,2}  &  I_{\eref{sunset-I145-1}} \\  \hline
	%
	%
	&   &   & \WH O_{0;I}  &    &    {\rm (I-2)} + {\rm (I-3)}\\  \hline
	&   &   &   &  \WH O_{0;I}  & {\rm (I-1)} - {\rm (I-3)}\\ % \hline
	%%
%	& \WH O_{0;2,1}  & -\WH O_{0;3,1}  &   &    &  I_{\eref{sunset-I234-m=0}} \\  \hline
	%%
%	-\WH O_{0;1,1} &   & \WH O_{0;3,1}  &   &    &  I_{\eref{sunset-I135-m=0}} \\  \hline
	%%
%	&   &   &   &    &  I_{\eref{sunset-I124plus125-m=0}}-I_{\eref{sunset-I234-m=0}}-I_{\eref{sunset-I135-m=0}} \\  
	%
\end{array}\right)~~~~~\label{sunset-massless-B-4}
\eea}
with
\bea \WH O_{0;I} & = &   =  6-3D-2\eta_4 {\d\over \d\eta_4}-2 \eta_5 {\d\over \d\eta_5} ~~~~~\label{sunset-massless-B-4-O}\eea
Solving them, we get $2$ T1B-type reduction rules for degree-1 operators ${\d\over \d\eta_a}, a=4,5$. From the fourth and fifth rows, we can find $2$ T2B-type reduction rules, i.e., using ${\d\over \d\eta_3}$ to solve ${\d\over \d\eta_i}, i=1,2$. 

{\bf Checking module of second round for massless case:} With new found reduction rules, the irreducible lattice points are $(0,0,n_3,0,0)$, thus we need to go to the third round of computation.

{\bf Third round computation of massless case:} Since only $n_3$ is not fully reduced, we can only act ${\d\over \d\eta_3}$ on $5$ reduction rules found in the second round of computation. The action on the reduction rule of ${\d\over \d\eta_4}$, after simplification, gives
\bea 0 &= & \left\{   \WH O_{0;II} {\d\over \d \eta_3} G_{111} \right\}+...~~~\label{sunset-m=0-Rn3}\eea
with
\bea \WH O_{0;II}=  { (D-4) K^2\over 2 s_0}  -{ K^2\over s_0}\eta_3 {\d \over \d \eta_3}~~~\label{sunset-m=0-Rn3-O}\eea
where $...$ is the part of lower degree and boundary.  	Using the result \eref{sunset-m=0-Rn3}, one can see that all nonzero $n_3$ can be reduced to $n_3=0$. So finally the only point can not be reduced using the reduction rule is the point $(0,0,0,0,0)$, which is the master integral for this case. 
Now we find the complete reduction rules.

{\bf Solving module of second round for massive case:} There are $6$ degree-2 equations with only $5$ of them to be independent. The coefficient matrix is 
{\tiny
\bea \left( \begin{array}{c |c |c |c|c|c|l} [1,1] &  [2,2] &[3,3] & [1,2] & [1,3]& [2,3] &  \\  \hline 
	& {-2m_2^2 \over s_0} &  & 	{-f_{+++}\over s_0} &  {-2m_3^2 \over s_0} &  {-2m_3^2 \over s_0} & {\rm (II-12)} \\  \hline 
	& & {-2m_3^2 \over s_0} & {-2m_2^2 \over s_0}	& {-f_{+++}\over s_0}    & {-2m_2^2 \over s_0}  & {\rm (II-13)} \\   \hline 
	{2m_1^2 \over s_0}& &  & 	{f_{+++}\over s_0} & {2m_3^2 \over s_0}  & {2m_3^2 \over s_0}  & {\rm (III-12)} \\ \hline 
	& & {2m_3^2 \over s_0} & {2m_1^2 \over s_0}	& {2m_1^2 \over s_0}& {f_{+++}\over s_0}    & {\rm (III-23)} \\ \hline 
	& {-2m_2^2 \over s_0} & {2m_3^2 \over s_0} & & & & {\rm (II-23)} \\ \hline 
	{2m_1^2 \over s_0} &  & {-2m_3^2 \over s_0} & & &  & {\rm (III-13)} 
\end{array}\right)~~~~~\label{sunset-massive-B-1}
\eea }
The coefficient matrix of $3$ degree-1 equations is 
{\tiny
\bea \left( \begin{array}{c| c |c |c| c| l} {\d\over \d\eta_1} & {\d\over \d\eta_2} &   {\d\over \d\eta_3} & {\d\over \d\eta_4} & {\d\over \d\eta_3} &  \\  \hline
	{2 m_1^2(m_2^2-m_3^2)\over s_0^2}  & {m_2^2 f_{++-}\over s_0^2} &  -{m_3^2 f_{+-+}\over s_0^2}   & -\WH O_{0;4,1}  &  \WH O_{0;5,2}  & {\rm (I-1)}  \\  \hline
	{m_1^2 f_{++-}\over s_0^2}	&  {2 m_2^2(m_1^2-m_3^2)\over s_0^2}  & -{m_3^2 f_{-++}\over s_0^2}   & \WH O_{0;4,2}  &  -\WH O_{0;5,1}   &   {\rm (I-2)} \\  \hline
	{m_1^2 f_{+-+}\over s_0^2}	  & -{m_2^2 f_{-++}\over s_0^2} & {2 m_3^2(m_1^2-m_2^2)\over s_0^2}  &  -\WH O_{0;4,1} &  \WH O_{0;5,1}  &  {\rm (I-3)}\\  
\end{array}\right)~~~~~\label{sunset-massive-B-3}
\eea}
Using \eref{sunset-massive-B-3}, we can solve ${\d\over \d\eta_a}, a=4,5$ by ${\d\over \d\eta_i}, i=1,2,3$. Using \eref{sunset-massive-B-1}, we can solve other $5$ degree-2 operators, using, for example, ${\d^2\over \d\eta_3^2}$.

{\bf Checking  module of second round for massive case:} Using reduction rules ${\d\over \d\eta_a}, a=4,5$, we can reduce $n_4, n_5$ to zero. Using five degree-2 reduction rules,
the points can not be reduced is the set $(0,0,n_3,0,0)$ and points  $(1,0,0,0,0)$, $(0,1,0,0,0)$. 
Again we need go to the third round computation. 

{\bf Third round computation of massive case:} Because only $n_3$ having not fully reduced, we just consider the action of ${\d\over \d\eta_3}$. Acting it on the reduction of, for example, the reduction rule of ${\d\over \d\eta_4}$, we can find the reduction rule of ${\d^2\over \d\eta_3^2}$.
Using this one, the un-reducible points are only following four, i.e.,
$(1,0,0,0,0)$, $(0,1,0,0,0)$, $(0,0,1,0,0)$ and $(0,0,0,0,0)$, which are exactly the master integrals. Thus we have found complete reduction rules.

%%%%%%%%%%%%%%%%%%%%%
\section{Example 2: The top sector of massless double box}
%%%%%%%%%%%%%%%%%%%%%%

For this one, four external momenta $k_1$, $k_2$, $k_3$, $k_4$ satisfy on-shell conditions $k_i^2=0$ and momentum conservation $\sum_{i=1}^4 k_i =0$, which leaves two independent scales $s=2 k_1 \cdot k_2$ and $t=2 k_2 \cdot k_3$. We choose a complete set of Lorentz scalars as
\begin{align}
	\begin{split}
		&\mathcal{D}_1={\ell_1}^2,\ \mathcal{D}_2=(\ell_1+k_1)^2,\ \mathcal{D}_3={(\ell_1+k_1+k_2)}^2,\\
		&\mathcal{D}_4= {\ell_2}^2,\ \mathcal{D}_5= {(\ell_2-\ell_1)}^2,\ \mathcal{D}_6={(\ell_2+k_1+k_2)}^2,\\
		&\mathcal{D}_7={(\ell_2-k_4)}^2,\ \mathcal{D}_8 = \ell_1 \cdot k_3,\ \mathcal{D}_9 = \ell_2 \cdot k_1,~~~\label{box-1-1-1}
	\end{split}
\end{align}
where the last two are ISPs.  Having presented the example of sunset with some details, we will be more briefly for this example. 

{\bf The first round computations:} The initial seed is $10$ DE's from IBP, where $2$ are degree-1  and $8$ degree-2. The combination of degree-2 DE's gives $2$ degree-1 DE's. So eventually we can find $4$ T1A-type reduction rules  for operators ${\d\over \d\eta_i}, i=1,3,4,6$. 
From the remaining $6$ degree-2 DE's, we can solve $6$ T1A-type reduction rules for ${\d\over \d\eta_a}{\d\over \d\eta_j}$, $a=8,9 $ and $j=2,5,7$.  
Using these $10$ reduction rules, the irreducible points are $(0,n_2,0,0,n_5,0,n_7,0,0)$ and $(0,0,0,0,0,0,0,n_8,n_9)$.

{\bf The second round computations:} Since the components $n_1, n_3, n_4, n_6$ have been fully reduced, we  consider the action  ${\d\over \d\eta_i}, i=2,5,7,8,9$ on above $10$  reduction rules. After simplification using descendant reduction rules, we will get $9$ new non-trivial degenerated equations, which can also be understood as the result of integrability conditions. 

With proper combinations, we have $4$ degree-2 DE's and $5$ degree-1 DE's. Using degree-2 DE's, we solve $3$ T1A-type reduction rules for degree-2 operators 
${\d^2\over \d\eta_2\d\eta_5}$, ${\d^2\over \d\eta_2\d\eta_7}$ and $ {\d^2\over \d\eta_5\d\eta_7}$ and  $1$ T1B-reduction rule for the operator ${\d^2\over \d\eta_8\d\eta_9}$.

For the $5$ degree-1 DE's, only $3$ of them contains good operators. Using them, we can solve $3$ T2B-reduction rules, i.e., using ${\d\over \d\eta_8}$ to solve ${\d\over \d\eta_9}$ and using ${\d\over \d\eta_5}$ to solve ${\d\over \d\eta_2}$ and ${\d\over \d\eta_7}$. 

With solved operators ${\d\over \d\eta_i}$, $i=2,7,8,9$, we can reduce $n_2, n_7, n_8, n_9$ to zero.  
Finally the irreducible lattice points are just
$(0,0,0,0,n_5,0,0,0,0)$ and $(0,0,0,0,0,0,0,1,0)$.

{\bf The Third round computations:} Now we need to act ${\d\over \d\eta_i}, i=8,5$ to new found reduction rules. Let us focus on the action of 
$5$ degree-1 DE's in previous round. Among them, we have degree-1 DE containing both ${\d\over \d\eta_8}$ and  ${\d\over \d\eta_5}$. We can use, for example, ${\d\over \d\eta_5}$ to solve ${\d\over \d\eta_8}$. We have also $2$ T1-type DE's. Using them we can solve $2$ T1B-type reduction rules for 
${\d^2\over \d\eta_8^2}$ and  ${\d^2\over \d\eta_5^2}$. Now we have reduce every $n_i$ to zero, except $n_5$ to $0,1$. Thus we get two master integrals, which is right for this sector. Now we find the complete reduction rules.

%%%%%%%%%%%%%%%%%%%%%%%%%%%%%%%%%%%%%%
%\section{The top sector of Two-loop massless non-planer}
%%%%%%%%%%%%%%%%%%%%%%%%%%%%%%%%%%%%%

%%%%%%%%%%%%%%%%%%%%%%%%%%%%%%%%%%%%%%
\section{Example 3: top sector of two-loop massless non-planer}
%%%%%%%%%%%%%%%%%%%%%%%%%%%%%%%%%%%%%

The kinematics are the same as planer double-box. We choose a complete set of Lorentz scalars as
\begin{align}
	\begin{split}
		&\mathcal{D}_1={\ell_1}^2,~~ \mathcal{D}_2=(\ell_1-k_1)^2,~~~\mathcal{D}_3={(\ell_1-k_1-k_2)}^2,\\
		&\mathcal{D}_4= {\ell_2}^2,~\mathcal{D}_5={(\ell_2+k_4)}^2,~\mathcal{D}_6= {(\ell_2-\ell_1+k_1+k_2+k_4)}^2,\\
		&\mathcal{D}_7={(\ell_2-\ell_1)}^2,~~~ \mathcal{D}_8 = \ell_1 \cdot k_4,~~~ \mathcal{D}_9 = \ell_2 \cdot k_1,~~~\label{nonplanerBox-1-1-1}
	\end{split}
\end{align}
where the last two are ISPs. 

{\bf The first round computation:} The initial seed is the $10$ DE's from fundamental IBP relations. From them, we can solve $2$ T1A-type reduction rules for degree-1 operators ${\d\over \d\eta_i}, i=1,3$, $6$ T1A-type reduction rules for degree-2 operators ${\d\over \d\eta_8}{\d\over \d\eta_j}, j=2,4,5,6,7$ and ${\d\over \d\eta_9}{\d\over \d\eta_2}$. There are also $2$ T2A-type reduction rules. One is the reduction rule of degree-2 operator
${\d\over \d\eta_9}{\d\over \d\eta_6}$, which depends on ${\d\over \d\eta_9}{\d\over \d\eta_7}$. Using these reduction rules,  irreducible lattice points are 
\bea {\cal U}_{11} &= &(0,n_2, 0,n_4, n_5, n_6, n_7, 0, 0),\nn
{\cal U}_{12} & = & (0,0, 0,0, n_5, 0, n_7, 0, n_9)\nn
{\cal U}_2 &= & (0,0, 0,0, 0, 0, 0, n_8, n_9) ~~~~~\label{np-un-point-2}\eea

{\bf The second round computation:} Since $n_1, n_3$ have been fully reduced, we should consider the action ${\d\over \d\eta_i}, i\neq 1,3$ on above $10$ reduction rules. Among $70$ equations, there are $16$ nontrivial DE's. Among them, there are $12$ degree-2 DE's and $4$ degree-1 DE's.  We can solve $10$ T1A-type reduction rules for degree-2 operators:  
  ${\d\over \d\eta_2}{\d\over \d\eta_4}$,  ${\d\over \d\eta_2}{\d\over \d\eta_5}$,  ${\d\over \d\eta_2}{\d\over \d\eta_6}$,  ${\d\over \d\eta_2}{\d\over \d\eta_7}$, ${\d\over \d\eta_4}{\d\over \d\eta_5}$, ${\d\over \d\eta_4}{\d\over \d\eta_7}$,  ${\d\over \d\eta_5}{\d\over \d\eta_9}$, ${\d\over \d\eta_5}{\d\over \d\eta_6}$,  ${\d\over \d\eta_6}{\d\over \d\eta_7}$,  ${\d\over \d\eta_7}{\d\over \d\eta_9}$. We get a T2A-type reduction rule, for, example, for 
  ${\d\over \d\eta_5}{\d\over \d\eta_7}$ by ${\d\over \d\eta_4}{\d\over \d\eta_6}$ and a T2B-reduction rule, for, example, for 
  ${\d\over \d\eta_8}{\d\over \d\eta_9}$ by ${\d^2\over \d\eta_8^2} $. From the degree-1 DE's, we can solve $3$ T2B-type reduction rules, for example, using ${\d\over \d\eta_4}$ to solve ${\d\over \d\eta_6}$, ${\d\over \d\eta_5}$ to solve ${\d\over \d\eta_7}$ and ${\d\over \d\eta_4}$ and ${\d\over \d\eta_5}$ to solve ${\d\over \d\eta_2}$.  Now the irreducible points are $(0,0,0,n_4,n_5,0,0,0,0)$, $(0,0,0,0,0,0,0,n_8,0)$ and $(0,0,0,0,0,0,0,0,n_9)$.

{\bf The third round computation:}  Since ${\d\over \d\eta_2}, {\d\over \d\eta_6}, {\d\over \d\eta_7}$ have been fully reduced, we consider the action of ${\d\over \d\eta_i}, i=4,5,8,9$ only. Also to generate less equations, we act only on  $3$ degree-1 reduction rules found in previous round ( If we can not find complete reduction rules, we can expand to degree-2 reduction rule late). Using these $12$ DE's, we can solve $3$ T1B-type reduction rules for degree-2 operators ${\d\over \d\eta_4}  {\d\over \d\eta_6}$, $ {\d^2\over \d\eta_4^2}$ and  ${\d^2\over \d\eta_5^2}$. Furthermore, we can solve ${\d\over \d \eta_9}$ by ${\d\over \d \eta_8}$ and ${\d\over \d \eta_5}$, then ${\d\over \d \eta_8}$ by ${\d\over \d \eta_4}$, and ${\d\over \d \eta_5}$ by ${\d\over \d \eta_4}$. Using there reduction rules, we can reduce $n_5, n_8, n_9$ to zero, and by the T1B-reduction rule of  ${\d^2\over \d\eta_4^2}$,  we can reduce $n_4$ to $0,1$, thus we get two master integrals for this sector.

\end{document}